\documentclass[preprint,12pt,authoryear]{elsarticle}

\usepackage{amssymb}
\usepackage{amsmath}

\usepackage{hyperref}
\usepackage{url}
\usepackage{natbib}
\graphicspath{{./}}
  
\usepackage{lineno}

\begin{document}

\begin{frontmatter}



\title{Probabilistic modelling of atmosphere-surface coupling with a copula Bayesian network} 


\author[uio]{Laura Mack\corref{cor}}
\author[met]{Marvin Kähnert}
\author[uio]{Norbert Pirk}

\affiliation[uio]{organization={Department of Geosciences, University of Oslo},
            city={Oslo},
            country={Norway}}
        
\affiliation[met]{organization={Development Centre for Weather Forecasting, Norwegian Meteorological Institute},
	city={Oslo},
	country={Norway}}

\cortext[cor]{Correspondence: laura.mack@geo.uio.no}

\begin{abstract} 
Land-atmosphere coupling is an important process for correctly modelling near-surface temperature profiles, but it involves various uncertainties due to subgrid-scale processes, such as turbulent fluxes or unresolved surface heterogeneities, suggesting a probabilistic modelling approach. We develop a copula Bayesian network (CBN) to interpolate temperature profiles, acting as alternative to T2m-diagnostics used in numerical weather prediction (NWP) systems. The new CBN results in (1) a reduction of the warm bias inherent to NWP predictions of wintertime stable boundary layers allowing cold temperature extremes to be better represented, and (2) consideration of uncertainty associated with subgrid-scale spatial variability. The use of CBNs combines the advantages of uncertainty propagation inherent to Bayesian networks with the ability to model complex dependence structures between random variables through copulas. By combining insights from copula modelling and information entropy, criteria for the applicability of CBNs in the further development of parameterizations in NWP models are derived.   
\end{abstract}




\end{frontmatter}


\section{Introduction}
Land-atmosphere coupling encompasses the turbulent exchange processes between the surface and the atmospheric boundary layer, and drives the energy and water cycle on various spatio-temporal scales, which also influences the occurrence of extreme events, such as heat waves, droughts or cold temperature extremes \cite[e.g.][]{Koster2005}. Since the land surface has a higher predictability than the atmosphere, it introduces a "land-surface memory", which leads to better predictability in numerical weather prediction (NWP) or climate models under strong surface-atmosphere coupling \citep{Ferguson2012}. Conversely, the decoupling of land and atmosphere cuts off the atmosphere from the surface information, thereby reducing its predictability. \\
Decoupling is a accompanying phenomenon of strongly stably stratified boundary layers (SBLs), that predominantly occur during calm clear-sky winter nights, when the turbulent heat flux is too small to counteract radiative cooling \citep{Derbyshire1998,vandeWiel2012}. While under coupled conditions, turbulence is continuous and mechanical mixing couples surface and boundary layer vertically, turbulence under decoupling is weak and intermittent, resulting in reduced vertical mixing and consequently colder near-surface temperatures. The turbulence under decoupling is affected by local instead of bulk shear, and buoyancy acts as sink for turbulent kinetic energy \citep{Zilitinkevich2013,Acevedo2016}. The strong stability can act as waveguide for submeso-scale motions, such as gravity waves or micro fronts, fading out the scale separation between turbulence and submeso-scale \citep{Mahrt2014}, ultimately leading to a breakdown of classical similarity theory \citep{Monin1954,Nieuwstadt1984} used to parameterize turbulence in NWP models. \\
In NWP models, these SBL-specific subgrid-scale processes are not represented, which manifests itself in the form of incorrectly modelled near-surface temperature profiles, with a particular inability to represent cold temperature extremes \citep{Day2024} and spatial variability \citep{Remes2025}. In mountainous terrain, the temperature bias depends additionally on physiographic features, usually with a pronounced warm bias in valleys and a cold bias at mountain peaks \citep{Rudisill2024}. This can be traced back partly to the model resolution and partly to the model physics. Increasing the model resolution leads to an improved representation of orographically-induced flows \citep{Wagner2014}. As a result, higher-resolution NWP models outperform coarser NWP models in the representation of typical stable boundary layer features, such as surface-based inversions and cold pools, even when the same model physics are used \citep{Mack2025}. However, incorrectly represented turbulent fluxes remain and entail problems in modelling the surface energy budget, especially over snow \citep{Gouttevin2023}, requiring improvements in the surface-layer turbulence parameterization itself. \\
Turbulence in NWP models consists of two main components: (1) the atmospheric turbulence, that connects atmospheric model levels, and (2) the surface fluxes, that couple the land surface to the atmospheric model. While atmospheric turbulence parameterizations \citep[e.g.][]{Lenderink2004}, usually perform well in single-column setups compared to large eddy simulations \cite[e.g.][]{Cuxart2006}, it is the surface fluxes that proved to be the decisive trigger for transitions between coupled and decoupled surface-atmosphere states, evidenced both by observations \citep{Acevedo2016} and NWP models \citep{Kaehnert2022,Kaehnert2025}. Overestimating vertical mixing (i.e. too strong coupling) leads to a warm bias \citep{Tjernstroem2021}, while underestimating vertical mixing (i.e. decoupling) leads to a cold bias \citep{Beljaars2011}. \\ 
When it comes to validation of the near-surface atmosphere, the most commonly used model output variable is the 2 m-temperature (T2m). However, unlike e.g. the surface temperature or the temperature on the first model level, T2m is a diagnostic and not a resolved model variable. In the here investigated NWP model HARMONIE-AROME \citep{Gleeson2024}, T2m is gained by interpolating surface and lowermost model level (at around 12~m) temperatures to the typical measurement height of 2~m following \citet{Geleyn1988}. This long-overlooked aspect has now gained considerable attention, as it has been shown that the T2m-diagnostic can have a significant impact on the results of model evaluation studies and can also be misused to d~eliberately influence temperature forecasts in the post-processing \citep{Preaux2025}. 
For example, \citet{Kaehnert2025} showed based on the NWP system HARMONIE-AROME that tuning the surface fluxes with a physically unreasonable shifted stability correction function can lead to an apparent improvement of 2 m-temperatures compared to observations, while the overall atmospheric temperature profile is represented worse due to the reduced turbulent surface fluxes.
Finding a consistent T2m-diagnostic that does not introduce an additional temperature bias is therefore considered as a key step that would subsequently also enable the further development of surface flux parameterizations \citep{Preaux2025}. \\
Here, we explore a new statistical approach to develop a probabilistic T2m-diagnostic. The aim is to diminish any stability-dependent temperature bias while representing uncertainty in land-atmosphere coupling and subgrid-scale spatial variability. This is achieved by providing a conditional probability density that represents a maximum-minimum range of possible temperatures. For this purpose, we use copula Bayesian networks (CBNs) -- a fusion of copulas and Bayesian networks \citep{Elidan2010}. 
Bayesian networks (BNs) use a graph structure to represent interrelations between variables \citep[e.g.][]{Berliner1998,Cano2004,Chen2012}. CBNs extend that framework by allowing for flexible quantile-dependent treatment of non-linear and non-Gaussian relationships through copulas. Copulas are particularly well suited for capturing multivariate dependence structures, making them useful in Earth system applications such as extreme events \citep[e.g.][]{Schoelzel2008,Tedesco2023} or hydrology \citep[e.g.][]{Salvadori2007}. In this study, we develop a CBN to predict T2m based on surface and atmospheric temperature, which we train based on observations from Finse, Norway \citep{Pirk2023,Mack2024}.  The CBN's behavior is evaluated in a synthetic sensitivity analysis, and subsequently applied to observations and for post-processing of HARMONIE-AROME model output. By combining copula modelling with information entropy, we further explore how copulas and CBNs can be applied to characterize land–atmosphere coupling and spatial thermal heterogeneity, aiming to derive criteria for their broader use in Earth system modelling.

\section{Data}

\subsection{The NWP system MEPS} \label{sec:nwp-param}
MEPS \citep{Frogner2019} is the regional NWP system for Scandinavia (Fig. \ref{fig:geo-data}) that builds on the HARMONIE-AROME configuration \citep{Gleeson2024}, and issues hourly forecasts up to 66 hours ahead on a grid with 2.5 km horizontal resolution and 65 terrain-following hybrid sigma levels (lowest model level at 12 m). 
The atmospheric turbulence is parameterized using the HARATU scheme \citep{Lenderink2004}, that describes turbulent fluxes between the atmospheric model levels using a 1.5-order TKE-closure, which performed well in an intercomparison with large eddy simulations \cite[e.g.][]{Cuxart2006}. The surface fluxes couple the lowermost atmospheric model level online to the surface model SURFEX \citep{Masson2013} using the bulk parameterization of \citet{Louis1979}. Due to a persistent warm bias in MEPS \citep{Koltzow2019} this schemes has been subjected to different adjustments in the used stability correction function which were tuned to improve the 2 m-temperature \citep{Kaehnert2025}. However, 2 m-temperature $T_{2\,m}$ is only a diagnostic (T2m) that is interpolated based on the prognostic (i.e. resolved) surface temperature $T_s$ and atmospheric temperature at the lowermost model level $T_a$ using the scheme from \citet{Geleyn1988}:
\begin{align} \label{eq:t2m-interpolation}
	T(z) &= T_s + s_i(z,z_0,z_{0H},Ri)\cdot(T_a-T_s).
\end{align}
The factor $s_i$ is a function of the requested interpolation height $z$ (with $z=2$ m for $T_{2\,m}$), the roughness length for momentum $z_0$, the roughness length for heat $z_{0H}$ and the bulk Richardson number
\begin{align}
	Ri := \frac{g}{T_0} \frac{\frac{\partial T}{\partial z}}{\left( \frac{\partial U}{\partial z} \right)^2},
\end{align}
describing dynamic stability as ratio of static stability and vertical wind speed ($U$) shear ($g$: gravity acceleration, $T_0$: reference temperature). Eddy diffusivity and conductivity are described using the neutral diffusivity scaled with a stability correction function for momentum and heat, respectively (\ref{appendix:most}).
Our study focuses on finding a new T2m-diagnostic that allows to calculate $T_{2\,m}$ from the resolved variables $T_s$ and $T_a$ only and represents uncertainty in the vertical coupling and subgrid-scale spatial variability. This would allow for (1) an improved representation of land-atmosphere coupling, (2) an uncertainty-aware model verification, and (3) omitting model tuning in the surface flux parameterizations. For this, we learn the statistical dependence structure (sec. \ref{sec:copulas}) based on observational data (sec. \ref{sec:obs-data}) and compare it to the \citet{Geleyn1988}-diagnostic in MEPS.

\subsection{Observations} \label{sec:obs-data}
\subsubsection{Flux tower observations at Finse}
We use data from two neighboring flux towers at Finse (60.11$^\circ$N, 7.53$^\circ$E, 1220 m a.s.l.) located in an alpine valley on the Hardangervidda mountain plateau in southern central Norway (Fig. \ref{fig:geo-data}).  At this seasonally snow-covered site, the surface layer is about 80 \% of the time stably stratified \citep{Pirk2023,Mack2024}, which makes it an ideal dataset to train data-driven models for stable boundary layers in complex terrain. \\
The main site Finse1 (Fig. \ref{fig:geo-data}c) is equipped with an eddy-covariance system at 4.4 m (CSAT3 three-dimensional sonic anemometer, Li-7200 closed-path infrared gas analyzer for measuring specific humidity $q$ and mixing ratio of CO$_2$ $c$), and provides temperature (2 m and 10 m), humidity (2 m and 10 m) and wind speed (10 m) measurements -- instrumentation is described in \citet{Pirk2023} and \citet{Mack2024}. Surface temperature (skin temperature measured with an infrared radiometer, SI-411, Apogee) is measured at approximately the same location as snow depth (laser distance sensor, SHM30, Jenoptik), which is used to correct the effectively changed measurement height of the fixed-installed in-situ sensors \citep{Preaux2025}.
The second site Finse2 (Finsefetene) is located 610 m southeast of the main site (along valley), and provides eddy-covariance measurements (same instruments as Finse1) and temperature measurements (2 m). Both eddy-covariance systems were sampled with a frequency of 10 Hz between February 2018 and January 2019. The post-processing with the R-package "Reddy" applies despiking, quality flagging and double-rotation, and outputs averaged quantities and fluxes on different integral time scales of 1, 2, 5, 10, 15, 30 and 60 minutes.

\subsubsection{NetAtmo observation network}
\begin{figure}
	\centering
	\includegraphics[width=13cm]{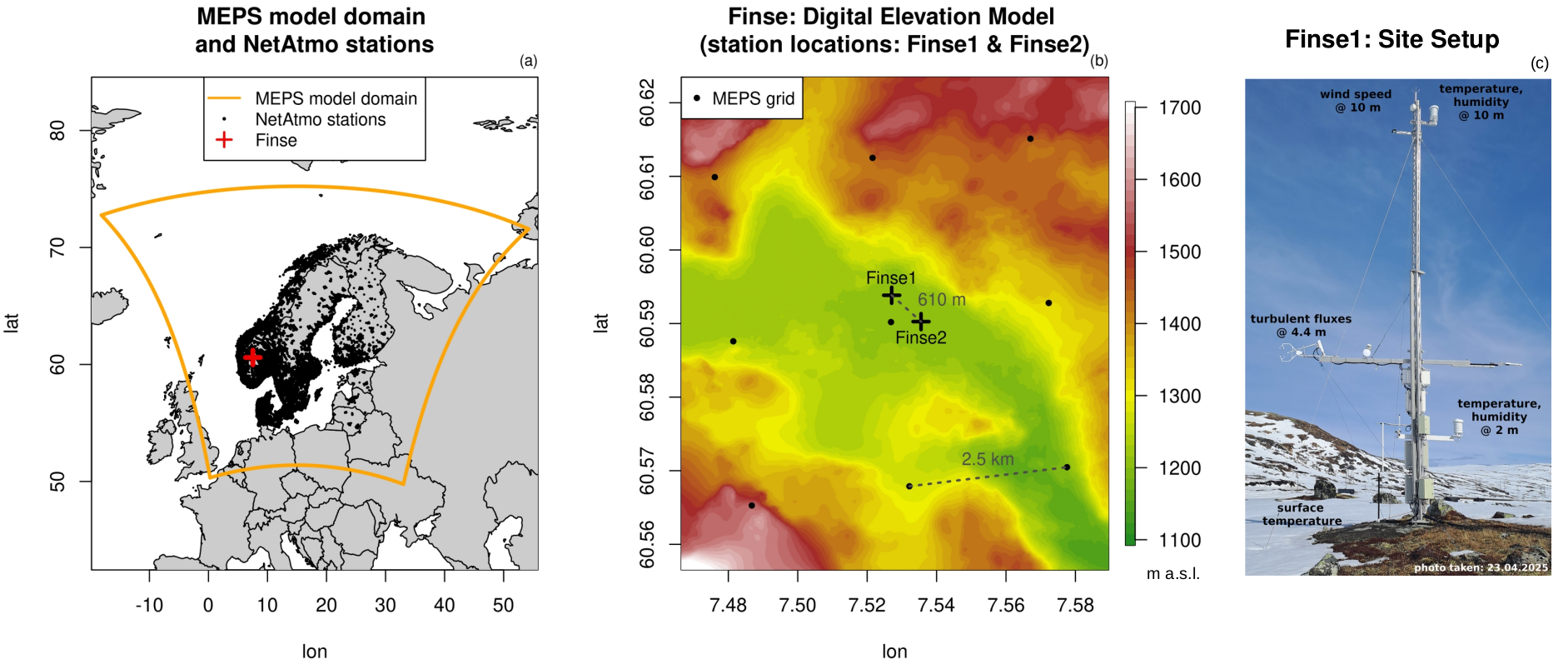}
	\caption{Used data sources: (a) Model domain of the NWP model MEPS and locations of the NetAtmo stations. (b) Surrounding of Finse and station locations. (c) Setup of the main site Finse1.}
	\label{fig:geo-data}
\end{figure}

In addition, temperature measurements from the NetAtmo network (Fig. \ref{fig:geo-data}a) subjected to a comprehensive quality control \citep{Baaserud2020} are used. 
Due to their high spatial coverage, these observations can capture local features that are typically unresolved in regional NWP models, such as cold pools and shallow temperature inversions that are particularly prevalent in stable boundary layers over complex topography. First studies point on advantages of using this network for post-processing of NWP model output \citep{Nipen2020}. Here, we use the hourly measurements of all Scandinavian NetAtmo stations that have the highest quality flag throughout a cold period in January 2024 (approximately 60 000 stations), allowing to characterize spatial temperature dependence over different terrain and static stabilities.

\section{Methods}

\subsection{Decoupling and thermal heterogeneity}
During surface layer decoupling, the turbulent heat flux vanishes resulting in a surface energy balance determined by radiative and ground heat fluxes only \citep{Derbyshire1998}. NWP models struggle to represent the bifurcation between coupled and decoupled state. This results in misrepresented near-surface temperatures in form of a warm bias when mixing is overestimated \citep{Tjernstroem2021}, or a cold bias when increasing static stability reinforces the decoupling of surface and first model level, referred to as "run-away" cooling \citep{Beljaars1998}, which is thus usually prevented in NWP models. 
However, a quasi-laminarization of the flow results in the same phenomenon in the real-world atmosphere \citep{Acevedo2016}, and then the flux calculated from eddy-covariance measurements does not represent the surface flux and has an unknown flux footprint. \\
To study decoupling based on observations, \citet{Peltola2021} introduced the decoupling metric $\Omega$ as dimensionless ratio between vertical velocity fluctuation $\sigma_w$ and critical vertical wind speed $w_{crit}$,  which describes the minimal wind speed a parcel needs to reach the ground against the hindering force caused by static stability, represented via the Brunt-Väisälä frequency $N^2 := g/T\, \partial T/\partial z$. $\Omega$ can be derived directly from the vertical momentum equation, i.e. without assuming (local) similarity theory, as
\begin{align}
	\Omega := \frac{\sigma_w}{\vert w_{crit} \vert} = \frac{\sigma_w}{\sqrt{2} z N}
\end{align}
and represents a thermodynamic measure for coupling without considering transport due to gravity waves.
For $\Omega > 1$, the vertical velocity fluctuation overcomes the static stability leading to vertical coupling, while for $\Omega < 1$, static stability suppresses the vertical velocity fluctuation resulting in decoupling. Since turbulent fluxes and vertical gradients can be strongly uncorrelated under stable decoupled conditions, considering  the turbulence intensity $\sigma_w$ directly (and not the turbulent flux) makes $\Omega$ superior to other stability measures in terms of distinguishing transport regimes under strong stratification \citep{Huss2024}. \\
Under strong stratification, turbulence is small-scale and can lead to pronounced horizontal heterogeneity \cite[e.g.][]{Mahrt2000}. 
While under strong large-scale forcing, surface heterogeneities are quickly blended with height and are thus negligible in NWP parameterizations \cite[e.g.][]{vanHeerwaarden2014}, they can affect mean-flow properties via dispersive fluxes under low wind conditions \citep{Margairaz2020a,Margairaz2020b}, explaining differences in temperature forecasts between NWP model output and observations \citep{Remes2025}. 
Here, we define thermal heterogeneity $H_T$ within an area as spatial standard deviation
\begin{align} \label{eq:thermal-heterogeneity}
	H_T := \sqrt{\langle T''^2 \rangle_{\textrm{area}}},
\end{align}
with $\langle ... \rangle$ representing spatial averaging and $''$ spatial perturbation. $H_T$ allows to quantify thermal heterogeneity (with the same unit as temperature) observation-based within grid cells corresponding to a NWP model setup.

\subsection{Statistical Methods} \label{sec:copulas}
\subsubsection{Copula theory}
Copulas are a flexible tool to construct multivariate joint distributions. Their key idea is to separate two tasks: estimating the individual behavior of each variable (the marginal distributions) and capturing how the variables depend on each other (the dependence structure described by a copula) \citep{Bardossy2008}. Copulas are invariant under monotonic transformations and describe the degree of dependence percentile-dependent. This means that the dependence structure can change for different parts of the distribution, e.g., extremes can have a different dependence structure than the median \citep{Bardossy2006}. In this regard, copulas are a generalization of scalar dependence measures, such as covariance or correlation (sec. \ref{sec:dependence-measures}). \\
The versatility of copulas has its roots in Sklar's theorem \cite{Sklar1959}, which states that the $d$-dimensional joint cumulative distribution function  $F(x_1,...,x_d) := \mathbb{P}(X_1\le x_1, ..., X_d\le x_d)$ can be established in terms of the copula $C:[0,1]^d \rightarrow [0,1]$ and the rank-transformed uniformly-distributed marginal distributions $F_1(x_1), ..., F_d(x_d)$ through
\begin{align}
		F(x_1,...,x_d) = C(F_1(x_1),...,F_d(x_d)).
\end{align}
If $F$ has the density function $f$, it can be decomposed into the copula density $c$ and the marginal densities $f_1,..., f_d$ through
\begin{align}\label{eq:sklar}
	f(x_1,...,x_d) = c(F_1(x_1), ..., F_d(x_d)) \cdot f_1(x_1) \cdot ... \cdot f_d(x_d).
\end{align}
The copula (density) is independent from the marginal distribution (density) and purely describes the dependence structure varying between the two limits of perfect negative and perfect positive association, referred to as Fréchet-Hoeffding bounds \citep{Frechet1951}. In the special case of independent variables, the joint distribution simply factorizes into the product of the marginals, with no copula effect. Thus, the copula density can be interpreted as a measure for coupling between two (or multiple) variables, whereby stronger coupling goes along with more certainty and higher predictability \citep{Ma2021}.

\subsubsection{Dependence measures and information entropy} \label{sec:dependence-measures}
Beyond copulas, dependence can also be captured through scalar measures. The well-known Pearson correlation coefficient $r(X_1,X_2) := \textrm{cor}(X_1,X_2)$ is limited to describing linear relationships and not invariant under monotonic transformations. Rank-based alternatives such as Spearman’s $\rho$ and Kendall’s $\tau$ avoid this issue, since they are derived directly from the copula.
Additional to these central measures of dependence, the upper and lower tail dependence coefficients ($\lambda_{u}, \lambda_{\ell}$) measure how likely high (low) values of a random variable $X_2$ given high (low) values of random variable $X_1$ are, i.e.
\begin{align}\label{eq:tail-dependence-coeff} 
	\lambda_{u} &:= \lim\limits_{t\rightarrow 1^-} \mathbb{P}(X_2>F_2^{-1}(t)\,\vert\,X_1>F_1^{-1}(t)), \\
	\lambda_{\ell}&:= \lim\limits_{t\rightarrow 0^+} \mathbb{P}(X_2\le F_2^{-1}(t)\,\vert\,X_1 \le F_1^{-1}(t)).
\end{align}
Different types of extremal dependence are distinguishing features of different copula families. \\
In analogy to the classical information entropy \citep{Shannon1948}, a copula entropy can be defined by applying the expected value $\mathbb{E}$ (with $\boldsymbol{u}:= (F_1(x_1), ..., F_d(x_d))$ as notation shortcut)
\begin{align}\label{eq:copent}
	H_C(\boldsymbol{u}) := - \int_{\boldsymbol{u}} c(\boldsymbol{u}) \ln(c(\boldsymbol{u})) \: d\boldsymbol{u} = - \mathbb{E}[\ln(c(\boldsymbol{u}))],
\end{align}
which corresponds to negative mutual information \citep{Ma2011} and is a non-linear multivariate measure of coupling uncertainty \citep{Ma2020}. 

\subsubsection{Copula modelling and pair-copula construction (PCC)} \label{sec:pcc}
Modelling joint densities or conditional densities in a copula framework consists of two steps \citep{Czado2022}: (1) Fitting the marginal distribution (or density), and (2) fitting the copula (density) based on the rank-transformed residuals. When both the marginal distributions and the copula are modelled parametrically, the approach is referred to as inference for margins (IFM). If the copula is modelled parametrically but the marginals empirically, the approach is called semi-parametric. 
To fulfill the requirement that the residuals are stationary and free of autocorrelation, \citet{Tootoonochi2021} provided a decision-support framework, which we follow here. The copula is estimated using maximum-likelihood estimation (MLE) and selected based on minimizing the Bayesian information criterion (BIC) using the copula families available in the R-package "VineCopula" \citep{VineCopula2025}. A goodness-of-fit (GOF) test with Kendall's process is performed \citep{Genest2006}. \\
Modelling multivariate copulas is limited due to the lack of generalizability of bivariate copula families to higher dimensions and computationally heavy. Pair-copula construction (PCC) \citep{Joe1996,Aas2009,Dissmann2013} overcomes these issues by representing multivariate densities iteratively through conditioning utilizing bivariate copulas only ("pair-copulas"), e.g., 
\begin{align}
	d=2:\quad \;& f(x_1 \vert x_2) = c_{12}(F_1(x_1),F_2(x_2)) \cdot f_1(x_1) \\
	d=3:\quad\;& f(x_1 \vert x_2, x_3) =  c_{13\vert 2}(F(x_1 \vert x_2), F(x_3\vert x_2)) \cdot f(x_1 \vert x_2), \label{eq:pcc3}
\end{align}
which can be iteratively extended to arbitrary dimensions $d$.
The involved marginal conditional density ("h-function") is given by the conditional copula
\begin{align}
	F(x_1 \vert x_2) = C_{1\vert 2}(F(x_1) \vert F(x_2)) = \frac{\partial C_{12}(F(x_1),F(x_2))}{\partial F(x_2)},
\end{align}
that is inferred via forward sampling using Rosenblatt transformation \citep{Rosenblatt1952}.
PCC for more than two variables contains a conditional copula that takes interrelation between the predictors (i.e. covariates) into account.
PCC lays the foundation for inference on complex tree and graph structures representing causal relations, and is used e.g. in geospatial interpolation \citep{Bardossy2008}, for vine copulas \citep{Czado2022} or copula Bayesian networks \citep{Elidan2010} (sec. \ref{sec:cbn-theo-background}).

\subsubsection{Copula Bayesian Networks (CBNs)} \label{sec:cbn-theo-background}
A Bayesian network (BN) is a directed acyclic graph that encodes dependencies with a conditional probability table, whereby nodes represent random variables and edges interrelations. Based on the graph structure, the conditional density can be decomposed in local terms that facilitate efficient inference and uncertainty propagation through the graph \citep{Chen2012}. However, classical BNs build on discrete random variables and cannot represent non-linear dependence structures \cite[e.g.][]{Hashemi2016}. \\
These drawbacks can be circumvented by using copula Bayesian networks (CBNs), a fusion of Bayesian networks and copulas \citep{Elidan2010}. CBNs build on the causal graph structure of BNs, but capture the dependence relations with copulas, i.e. the conditional density is re-parametrized with copulas, inheriting the properties of copulas to capture general non-linear dependencies of continuous random variables. A very efficient re-parameterization for this purpose is PCC (sec. \ref{sec:pcc}) due to the use of local bivariate copulas as building blocks \citep{Bauer2016}. Translated to Bayesian language, the marginal density $f(x_1)$ in the PCC (eq. \ref{eq:pcc3}) represents the prior density and the resulting conditional density $f(x_1 \vert x_2, x_3)$ the posterior density.
The CBN modelling procedure consists of three steps: 
\begin{enumerate}
	\item[(1)] graph structure learning: identify nodes and learn graph structure (as for classical BNs)
	\item[(2)] copula learning: fitting marginal densities and copulas according to PCC
	\item[(3)] model use (inference)
\end{enumerate}
In our analysis, we construct a simple graph structure (step 1) to infer $T_{2\,m}$ from $T_s$ and $T_a$ as predictors (parent nodes of $T_{2\,m}$) similar to the existing T2m-diagnostic in NWP models (sec. \ref{sec:nwp-param}). We apply PCC according to eq. \ref{eq:pcc3} and fit the required copulas parametrically based on observational data (step 2). For the marginal densities, we compare parametric and non-parametric approaches. The CBN is applied to a synthetic example in form of a sensitivity analysis, to tower observations and for post-processing of MEPS output (step 3).

\section{Results}

First, the relation between vertical coupling and spatial heterogeneity is analyzed utilizing copula entropy, before proceeding with copula modelling and construction of a CBN, whose properties and applications are then explored.

\subsection{Relation of vertical and horizontal coupling} \label{sec:results-hor-vert-coupling}
\begin{figure}
	\centering
	\includegraphics[width=12cm]{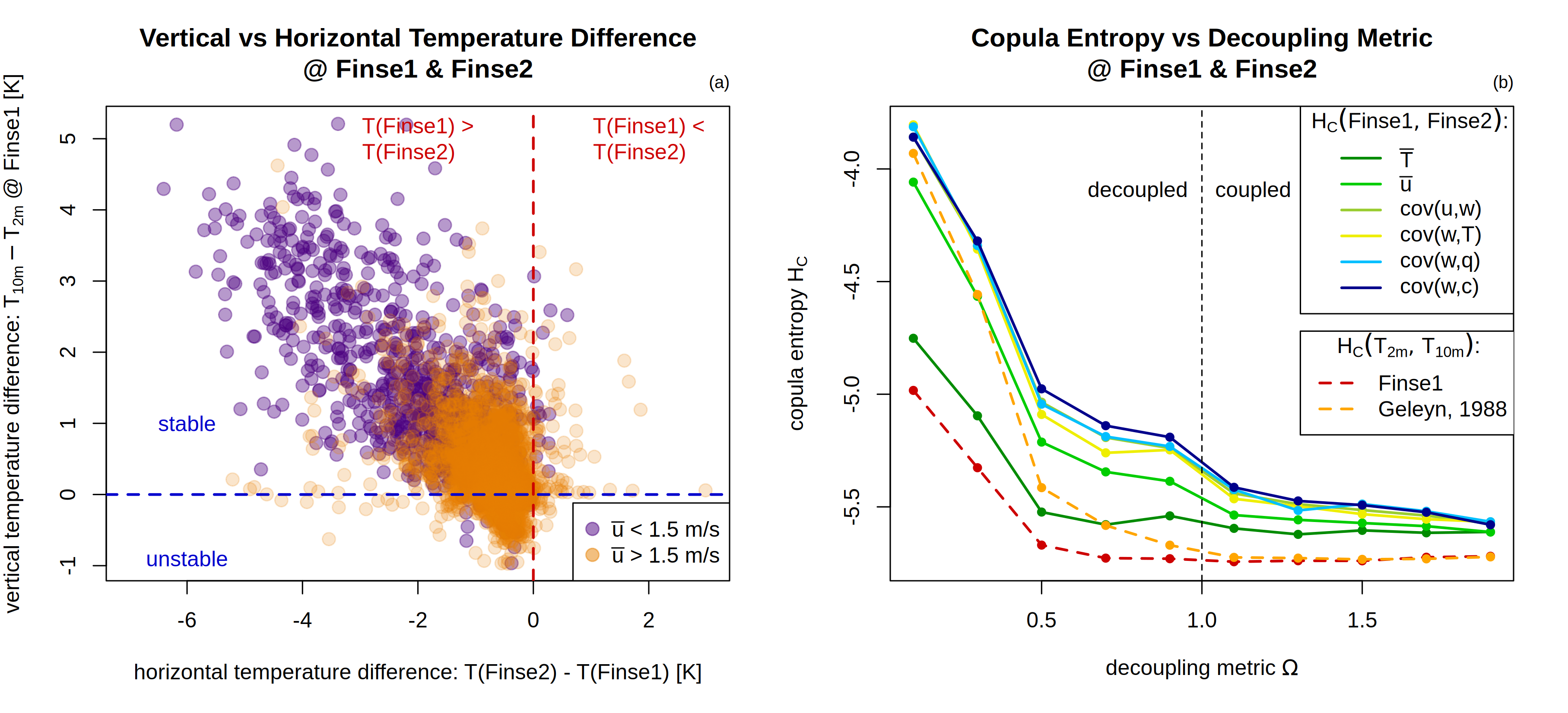}
	\caption{Relation of vertical and horizontal coupling: (a) Horizontal temperature difference between Finse1 and Finse2 versus vertical temperature difference at Finse1 color-coded based on a wind speed threshold (30 minutes temporal average), (b) decoupling metric at Finse1 versus copula entropy between Finse1 and Finse2 of temperature, wind speed, and the covariances corresponding to the fluxes of momentum cov(u,w), sensible heat cov(w,T), latent heat cov(q,w) and CO$_2$ cov(c,w) (30 minutes temporal average). Entropy calculation is based on time series of the same length to avoid a sampling size bias.}
	\label{fig:hor-vs-vert-coupling}
\end{figure}

Fig. \ref{fig:hor-vs-vert-coupling}a investigates the relation of horizontal and vertical coupling exemplary based on the two neighboring sites at Finse (sec. \ref{sec:obs-data}). The horizontal temperature difference between the two sites increases with increasing vertical temperature difference, reaching a maximum of up to 6 K, corresponding to an average horizontal gradient of about 1 K per 100 m, which is in the order of magnitude of the dry adiabatic lapse rate. These large horizontal temperature differences occur almost exclusively under low wind conditions (below 1.5 m/s), confirming that pronounced thermal heterogeneity occurs under strong stability and high Richardson number. \\ 
Fig. \ref{fig:hor-vs-vert-coupling}b shows the copula entropy of mean temperature, wind speed and fluxes (solid lines) between the two sites at Finse in dependence on the decoupling metric $\Omega$. With increasing vertical decoupling (low $\Omega$ values), the copula entropy increases, indicating increased information loss between the two stations, i.e. the targeted variables evolve more independently. Hereby, the information loss is higher for the fluxes than the first-order statistics. The copula entropy between $T_{2\,m}$ and $T_{10\,m}$ (dashed lines) is shown for the actual measurements (red) and the  T2m-diagnostic from \citet{Geleyn1988} (orange), which is calculated by inserting the observations into eq. \ref{eq:t2m-interpolation}. Both increase with decreasing $\Omega$, indicating that physical decoupling (in terms of $\Omega$) goes along with vertical thermal information loss. The increase is stronger for the parametrized case, suggesting that the T2m-diagnostic is not making efficient use of the available information. 
In summary, physical processes that drive uncertainty in stable boundary layers, such as vertical decoupling and horizontal heterogeneity, translate into the world of information theory by an increased loss of information, both in vertical and horizontal direction. This reflects that nearby measurements become more independent, which can be interpreted as 'information decoupling'. In NWP models, these processes are sub-grid scale, so copula entropy provides a valuable framework to quantify and diagnose the uncertainty in these unresolved processes.

\subsection{Modelling dependence structures using bivariate copulas} \label{sec:results-copulas}
\begin{figure}
	\centering
	\includegraphics[width=9cm]{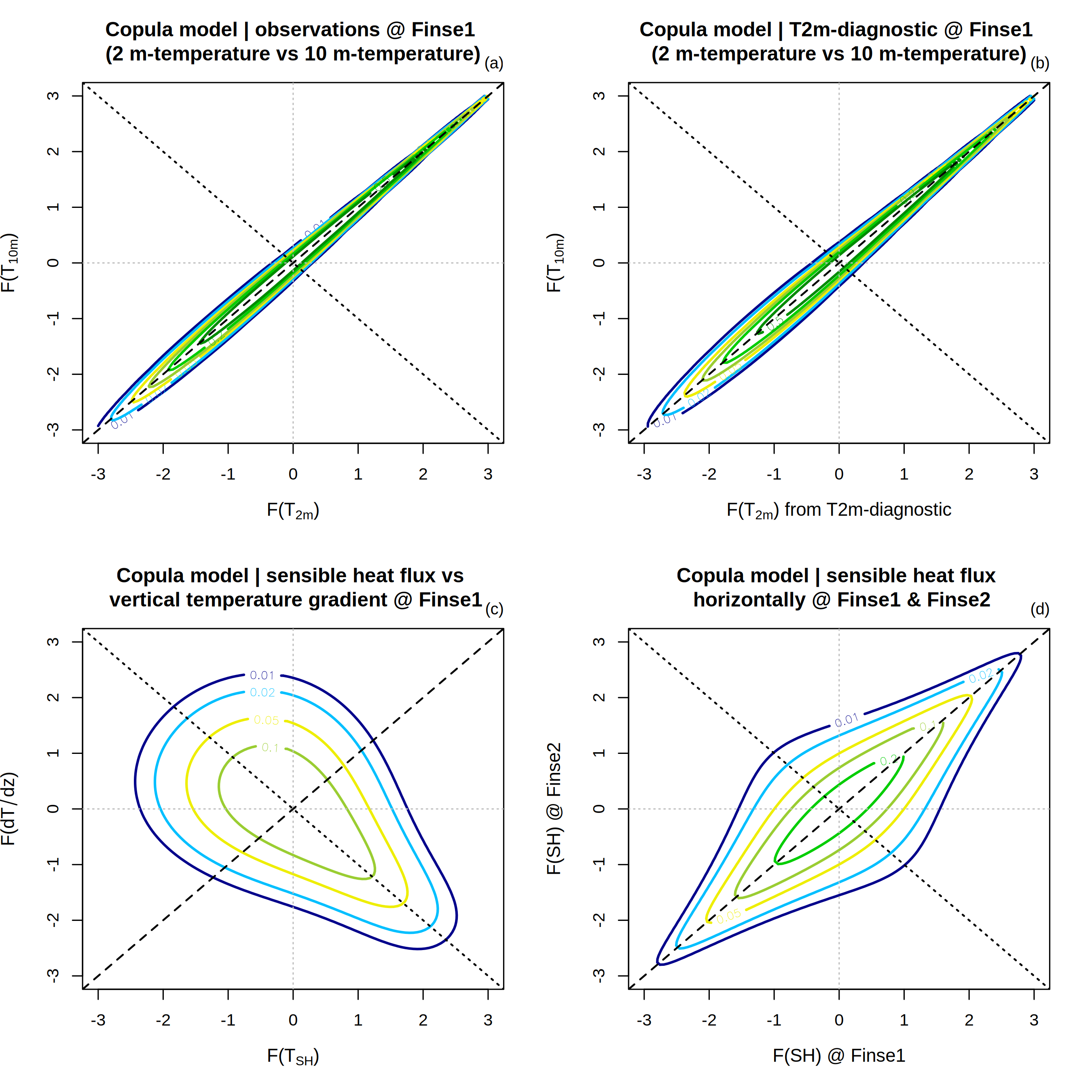}
	\caption{Examples of fitted bivariate copula models for the steady residuals of  (a) $T_{2\,m}$ vs $T_{10\,m}$ based on Finse1, (b) $T_{2\,m}$-diagnostic interpolated using eq. \ref{eq:t2m-interpolation} vs $T_{10\,m}$ based on Finse1,
		(c) sensible heat flux vs vertical temperature gradient, (d) sensible heat flux at Finse1 vs Finse2. Plotted w.r.t. normalized margins. First diagonal (dashed), second diagonal (dotted).}
	\label{fig:copula-examples}
\end{figure}

In the next step, we move on to modelling the dependence structure with copulas (sec. \ref{sec:copulas}). Tab. \ref{tab:copulas-model-selection} summarizes the bivariate copulas that are selected based on minimizing BIC and Fig. \ref{fig:copula-examples} showcases some of the selected copulas. \\
The dependence relation between  $T_{2\,m}$ and $T_{10\,m}$ follows for both measured (Fig. \ref{fig:copula-examples}a) and parametrized $T_{2\,m}$ (Fig. \ref{fig:copula-examples}b) a narrow Tawn type2 copula. This copula type is characterized by an asymmetric tail dependence with stronger upper than lower tail dependence, indicating a stronger dependence for positive temperature residuals. This means that warmer-than-expected conditions (positive residuals) are more strongly coupled across heights than colder-than-expected ones (negative residuals). In the parameterized case, the lower tail dependence is smaller, suggesting greater uncertainty in how cold anomalies are handled. A similar pattern can also be observed when comparing the coupling between surface and atmospheric temperature in the observations and MEPS (closest grid point to Finse, not shown). \\
The dependence between sensible heat flux (SH) and the vertical temperature gradient (dT/dz) (Fig. \ref{fig:copula-examples}c) is more uncertain, indicated by the wider copula that relates to higher copula entropy. The dependence between negative residuals of dT/dz and positive residuals of SH (unstable stratification) is stronger, and thus less uncertain, compared to the relation of positive residuals of dT/dz and negative of SH (stable stratification). This confirms that under stable stratification the temperature gradient is a worse predictor of SH than under unstable, caused by the high amount of counter-gradient SH found at this station \citep{Mack2024}. \\
When modelling the dependence between Finse1 and Finse2, mostly the t-copula is selected, as exemplified in Fig. \ref{fig:copula-examples}d for sensible heat flux. However, for latent heat flux and momentum flux a strong asymmetry both w.r.t. the first and second diagonal can be observed (based on the copula type listed in Tab. \ref{tab:copulas-model-selection}). For latent heat flux the upper tail dependence is much stronger than lower tail dependence, indicating that evaporation occurs more often simultaneously at both stations while the relation of condensation/deposition between the stations is more uncertain. For momentum flux, a strong asymmetry w.r.t. the first diagonal can be found, which indicates an asymmetric information content between the two stations probably related to effects of micro-topography. For shorter averaging time scales, the copulas become wider, indicating decreasing correlation between the two sites. \\
Exploring the shapes of these bivariate copulas, helps to assess the potential of constructing a CBN with PCC to predict them. The wider the copula is, the higher the copula entropy and the more uncertain the coupling between the considered variables, implying that predicting one variable with the other becomes more uncertain. An asymmetry w.r.t. the second diagonal indicates different tail dependence, while an asymmetry w.r.t. the first diagonal indicates systematic differences between the variables or in case of spatial variables a spatial asymmetry, that relates to a dependence on location. 

\begin{table}
	\footnotesize
	\centering
	\begin{tabular}{l|llcrr}
		\hline
		\textbf{residuals}& \textbf{data} &\textbf{selected copula} &\textbf{ parameter(s)} &  \textbf{BIC} & \textbf{logLik} \\ 
		\hline
		$T_s, \, T_{12\, m}$ & MEPS & BB8 & (6, 0.96)  & -3160 & 1588 \\
		$T_s, \, T_{10\, m}$ & Finse1 & Gaussian & 0.96 & -17 369 & 8 689 \\		
		$T_{2\,m}, \, T_{10\, m}$ & Finse1 & Tawn type 2 & (20, 1) & -35696 & 17857\\
		$T_{2\,m}, \, T_{10\, m}$ & Finse1/eq. \ref{eq:t2m-interpolation} & Tawn type 2 & (14.7, 1) & -13482 & 6749\\
		\hline
		$dT/dz, U$ & Finse1 & BB8 270$^\circ$-rot.& (-1.8, -1)& -1430 & 723\\
		$SH, dT/dz$ & Finse1 &BB8 270$^\circ$-rot.& (-2, -1) & -1983 & 1000\\
		\hline
		$T$ & Finse1\&2 & student-t & (1, 30) & -20183 & 10100\\
		$U$& Finse1\&2 & student-t & (0.9, 2.7) & -10492 & 5255 \\
		cov$(w,T)$& Finse1\&2 & student-t & (0.77, 2) & -4908 & 2463  \\
		cov$(w,q)$& Finse1\&2 & BB8 & (2.8, 1) & -3825 & 1921 \\
		cov$(w,c)$& Finse1\&2 & student-t & (0.56, 2) & -2929 & 1473 \\
		cov$(w,u)$& Finse1\&2 & Tawn type 1 180$^\circ$-rot. & (2.7, 0.7) & -5510 & 2764\\
		\hline
	\end{tabular}
	\caption{Overview of the selected bivariate copulas: Pairs of considered quantities, data source, selected copula (based on minimizing BIC), fitted copula parameter(s), Bayesian information criterion (BIC), and log-likelihood (logLik).}
	\label{tab:copulas-model-selection}
\end{table}

\subsection{Probabilistic modelling of near-surface temperature profiles using a CBN} \label{sec:results-validation-on-obs}
\begin{figure}
	\centering
	\includegraphics[width=12cm]{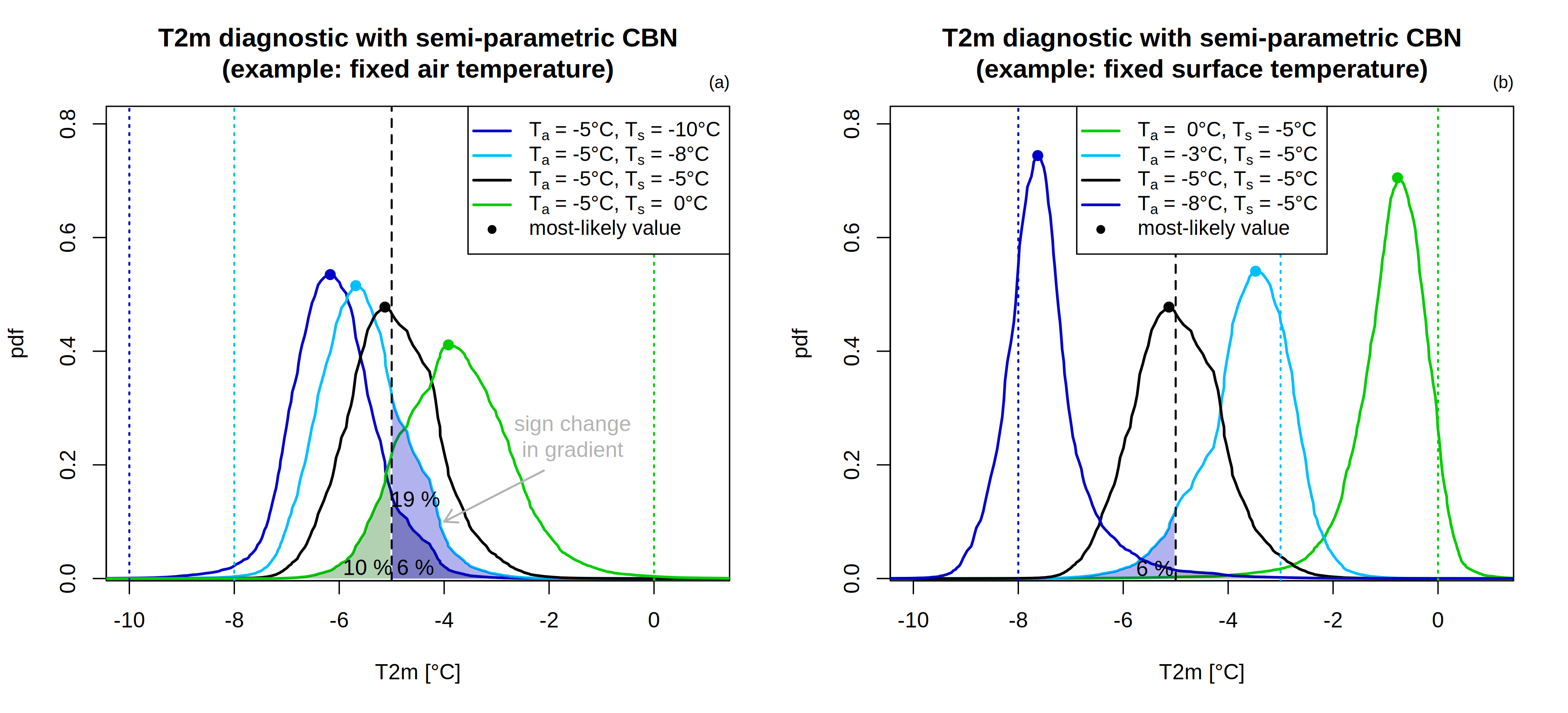}
	\caption{Conditional (i.e. posterior) density function $f(T_{2\,m} \vert T_s, T_a)$ resulting from CBN inference: (a) Example with fixed 10 m-atmospheric temperature of $T_a = -5^\circ C$ and varying surface temperatures (colored). (b) Example with fixed surface temperature of $T_s = -5^\circ C$ and varying atmospheric temperatures (colored). The shaded areas represent the probability for a sign change in gradient below measurement height, and the dot the most-likely value of the respective density. }
	\label{fig:t2m-cbn-probabilistic}
\end{figure}

\begin{figure}
	\centering
	\includegraphics[width=12cm]{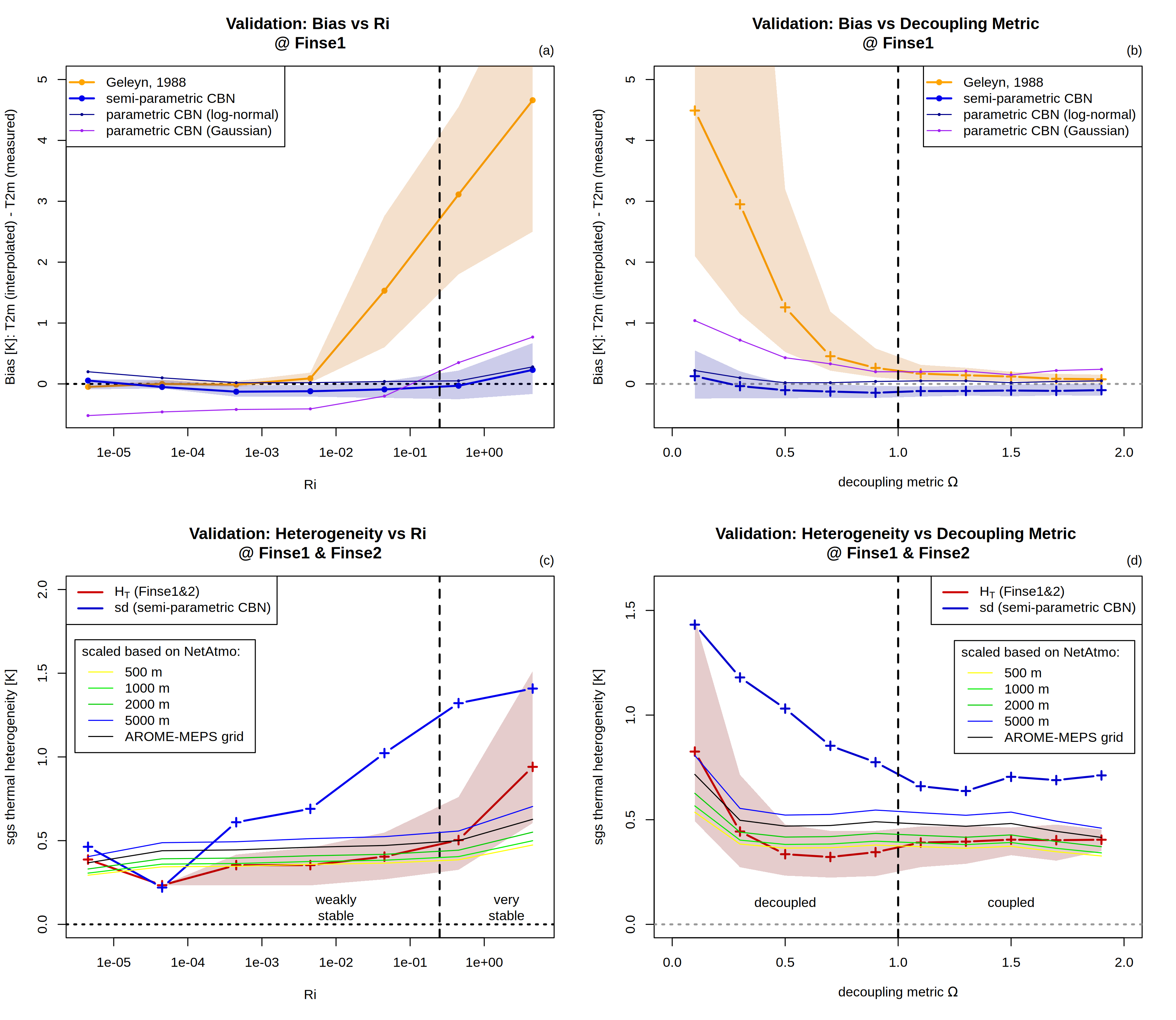}
	\caption{Validation of the interpolated $T_{2\,m}$ from the CBN compared to the \citet{Geleyn1988}-parameterization and observations: Bias of the predicted $T_{2\,m}$ from the CBN (blue) and \citet{Geleyn1988}-parameterization (orange) w.r.t. to the observations at Finse1 based on Richardson number (a) and decoupling metric (b). Standard deviation of $f(T_{2\,m} \vert T_s, T_a)$ from the CBN (blue), estimate for subgrid-scale thermal heterogeneity utilizing the most-likely value of $T_{2\,m}$ from the CBN as input for the scaling learned from the NetAtmo data for different grid resolutions (colored thin lines), $H_T$ between Finse1 and Finse2 (red), based on Richardson number (c) and decoupling metric (d).}
	\label{fig:t2m-cbn-evaluation}
\end{figure}

Sec. \ref{sec:copulas} showed that the vertical coupling of the temperature measurements can be modelled with peaked copulas (low copula entropy), suggesting the potential of creating a CBN to predict them precisely. Here, we construct and train based on Finse1 a CBN with PCC according to eq. \ref{eq:pcc3} with $T_s$ and $T_a$ as predictor for $T_{2\,m}$, which involves the bivariate copula ($T_{2\,m},T_a$) shown before in Fig. \ref{fig:copula-examples}a, and a pair-copula representing predictor interrelation. $T_s$ and $T_a$ have the highest rank correlation with $T_{2\,m}$, i.e. they are selected in score-based graph structure learning using Kendall's $\tau$. 
The output of the CBN is the conditional probability density function $f(T_{2\,m} \vert T_s, T_a)$, that describes the uncertainty of $T_{2\,m}$ given $T_s$ and $T_a$. \\
Fig. \ref{fig:t2m-cbn-probabilistic} shows $f(T_{2\,m} \vert T_s, T_a)$ calculated based on the semi-parametric CBN (parametric copulas, non-parametric marginal densities). We conduct two sensitivity analyses with (a) fixed 10 m-atmospheric temperature $T_a$ and (b) fixed surface temperature $T_s$. In all cases, the most-likely value of the resulting conditional density lies between the surface and atmospheric temperature. For some conditions, however, there is a probability for a sign change in the vertical temperature gradient below the measurement height (indicated by the shaded area), which corresponds either to a shallow surface-based inversion (i.e. maximum temperature below measurement height), or a shallow unstable layer (i.e. minimum temperature below measurement height). Representing these unresolved features in the temperature profile below first model level is in current NWP models not possible. 
From comparing the two subplots (Fig. \ref{fig:t2m-cbn-probabilistic}a,b) it can be seen, that $T_{2\,m}$ is more strongly coupled to the atmospheric temperature $T_{10\,m}$, than to the surface temperature $T_{s}$, which is in agreement with the selected copulas (Tab. \ref{tab:copulas-model-selection}). \\
Fig. \ref{fig:t2m-cbn-evaluation} shows an out-of-sample validation of the trained CBN compared to the (deterministic) \citet{Geleyn1988}-diagnostic based on Finse1 depending on Richardson number $Ri$ and decoupling metric $\Omega$. Different marginal densities were tested to translate the univariate residuals back to temperature. The semi-parametric CBN uses, as before, a kernel density estimate of $T_{2\,m}$, while the parametric CBN explores both a Gaussian and a scaled log-normal fit. \\
The T2m-diagnostic of \citet{Geleyn1988} shows a strong warm bias with increasing $Ri$ and decreasing $\Omega$, which reaches more than 4 K under strongly stable and decoupled conditions (Fig. \ref{fig:t2m-cbn-evaluation}a,b). In contrast, the semi-parametric CBN and the log-normal parametric CBN (using the most-likely value) do not show a bias, and the $T_{2\,m}$-prediction accuracy is independent of stratification and coupling state. The parametric CBN with Gaussian marginal density shows a cold bias for weakly stable coupled situations and a slight warm bias under stable decoupled conditions, which is due to the misrepresentation of the tail structure of $T_{2\,m}$ with a Gaussian density. This indicates that additionally to the copula learning also the accurate fitting of the marginal density, particularly its tail structure, is of importance. \\
Fig. \ref{fig:t2m-cbn-evaluation}c,d investigate the variance of $f(T_{2\,m} \vert T_s, T_a)$. The standard deviation of $f(T_{2\,m} \vert T_s, T_a)$ increases with increasing $Ri$ and decreasing $\Omega$, indicating that the uncertainty of the interpolated $T_{2\,m}$ is higher under stable and decoupled conditions. Spatial variability is also captured: the mean absolute difference in $T_{2\,m}$ between Finse1 and Finse2 increases with stability and decoupling (Fig.\,\ref{fig:t2m-cbn-evaluation}c,d). \\
Finally, the most-likely value of $T_{2\,m}$ from the semi-parametric CBN was used to estimate the subgrid-scale spatial thermal heterogeneity $H_T$ utilizing the found linear relation between $\langle T \rangle$ and $H_T$ based on the NetAtmo data (\ref{appendix:scaling-heterogeneity}). The resulting thermal heterogeneity increases systematically with larger area, and also for high $Ri$ and low $\Omega$. Although the approaches are not directly comparable, this provides a physically meaningful interpretation, that, in stable or decoupled conditions, the CBN-predicted uncertainty grows in a manner consistent with increased subgrid-scale thermal heterogeneity.

\subsection{Application in NWP model output post-processing}
\begin{figure}
	\centering
	\includegraphics[width=14cm]{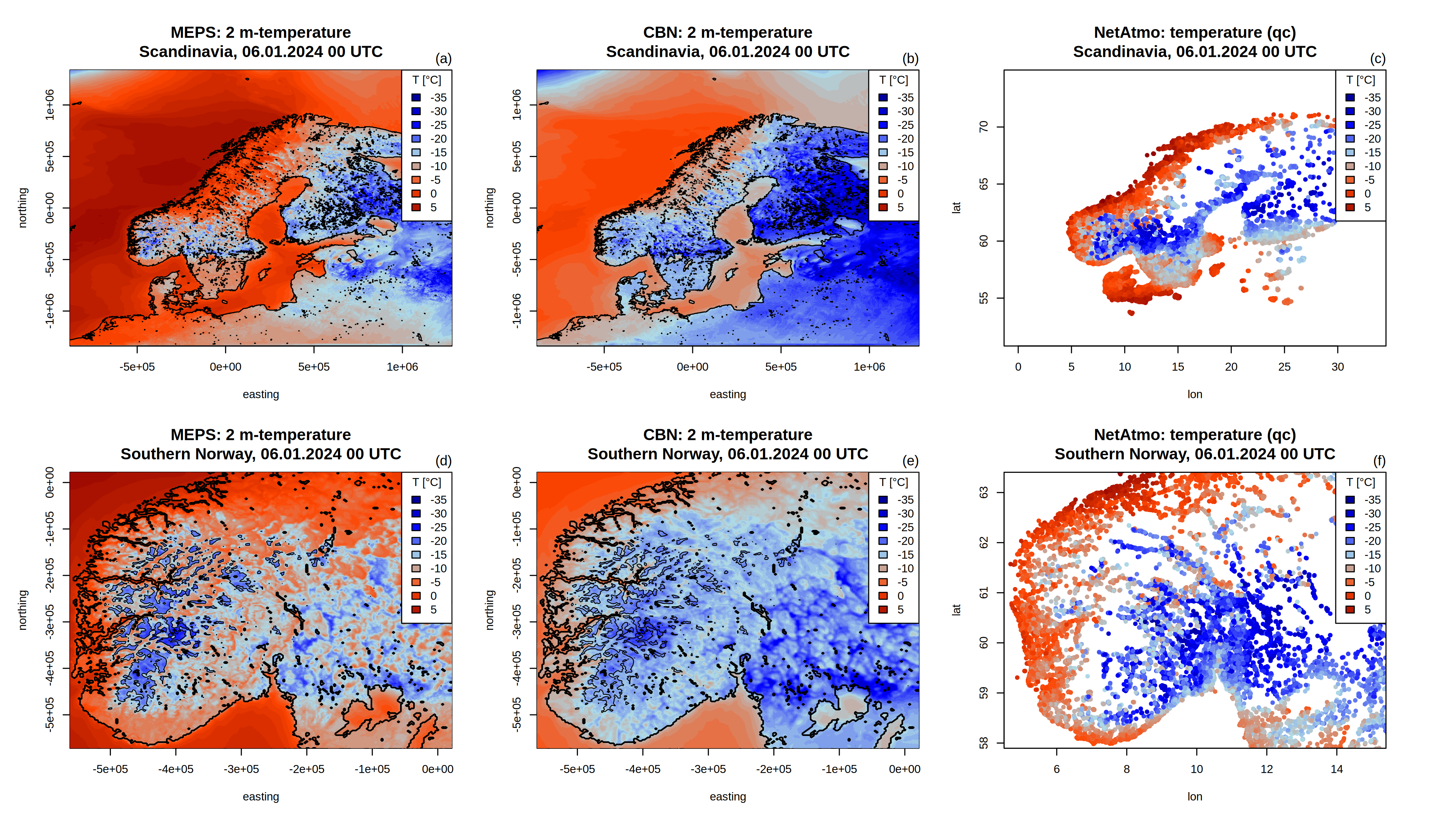}
	\caption{CBN applied to MEPS in a case study from 06.01.2024 00 UTC: MEPS T2m-diagnostic (a, d), T2m-CBN (b, e) and NetAtmo observations (c, f) for the whole model domain (upper row a-c) and zoomed-in for southern Norway (lower row d-f).}
	\label{fig:case-study-cbn-applied-to-meps}
\end{figure}

In the last step, we explore the applicability of the trained CBN to post-process MEPS model output. As case study a period of particularly cold temperatures over Scandinavia in January 2024 caused by a high pressure system centered over southeast Norway is explored. We use the resolved model fields of $T_s$ and $T_a$ from MEPS as input for the CBN to predict $T_{2\,m}$. The prior density (marginal density of $T_{2\,m}$) is constructed parametrically using a scaled log-normal density (see sec. \ref{sec:results-validation-on-obs}) fitted to the NetAtmo observations at the previous time step, mimicking an operational near-real-time application. A weighting function takes the difference in the height of the atmospheric temperature from the training data (10 m) to the model data (12 m) into account. \\
Compared to the gridded NetAtmo observations, the T2m-CBN (bias: -0.4 K, RMSE: 6.3 K) outperforms the T2m-diagnostic from MEPS (bias: 5.9 K, RMSE: 8.3 K) and also shows a lower copula entropy, indicating higher accuracy in the predicted 2 m-temperature, despite given the same input fields.
Fig. \ref{fig:case-study-cbn-applied-to-meps} shows the $T_{2\,m}$ fields from the MEPS T2m-diagnostic and the T2m-CBN for the whole model domain and zoomed-in for southern Norway as well as the respective NetAtmo observations, exemplary for the time-step 06.01.2025 00 UTC. Thereby, the MEPS T2m-diagnostic shows higher temperatures than the T2m-CBN, except in some valleys in the Norwegian mountains. The behavior of the MEPS T2m-diagnostic corresponds to the findings of previous studies \citep{Koltzow2019,Kaehnert2025}, where the typical warm bias can turn into a cold bias as result of the tuning in the flux parameterization, which weakens the atmospheric inversion strength when land and atmosphere decouple. In contrast, the T2m-CBN captures the cold temperature extremes in southern Norway well, but exhibits a cold bias along the coast, which is not surprising given that such data was not included in the training data. 
In summary, despite the same input fields, the T2m-diagnostic has a strong effect on the interpolation of near-surface temperature, whereby the T2m-CBN greatly reduces the warm bias but without inducing unnatural decoupling.

\section{Discussion}

\subsection{Decoupling, spatial heterogeneity and copula entropy}
The observation-based analysis (sec. \ref{sec:results-hor-vert-coupling}) shows that spatial thermal heterogeneity increases with increasing stability and increasing land-atmosphere decoupling. This spatial heterogeneity is typically a subgrid-scale feature and thus unresolved in NWP models, which complicates model verification based on point observations \citep{Goeber2008}, especially in the case of strong stratification \citep{Remes2025}. \\
Spatial heterogeneity can be interpreted as horizontal decoupling, going along with a loss of information between neighboring measurement stations, which motivates the translation of the concept of 'physical decoupling' to 'information decoupling' in the context of information theory. For this purpose, we transfer the theoretical concept of copula entropy \citep{Ma2011,Ma2020} to applications in meteorological measurement setups, ultimately allowing to objectively quantify coupling uncertainty within a spatially distributed measurement setup. 
Using copula entropy as measure for coupling uncertainty is a novel concept, whereas classical information entropy is a well established measure for flow organization \citep[e.g.][]{Wesson2003}. In organized flows, which are characterized by eddies of similar size and a peaked energy spectrum, information entropy is low compared to disorganized flows, where interacting eddies of various sizes lead to a flat energy spectrum \citep{Fu2014}. \\
Copula entropy can also be used to compare the information loss between model and observations, both horizontally and vertically. It is shown that the T2m-diagnostic from \citet{Geleyn1988} used in HARMONIE-AROME leads to an increased information loss between surface and near-surface atmospheric temperature compared to observations. In this regard, copula entropy could also enrich model evaluation and benchmarking studies as a new verification metric.

\subsection{Probabilistic T2m-diagnostic -- advancements \& limitations} 
While the use of BNs for parameterizing subgrid-scale processes has already been proposed \citep{Frederiksen2008,OKane2024}, CBNs have not yet been applied for this purpose despite their greater flexibility. Overall, the performance of the T2m-CBN provides proof-of-concept for further applications of CBNs in Earth system modelling. \\ 
The here new developed T2m-CBN takes surface and atmospheric temperature as input to predict the entire conditional density of 2 m-temperature. Applying the T2m-CBN to observations and NWP model output indicates several advancements compared to the currently used T2m-diagnostic from \citet{Geleyn1988} in HARMONIE-AROME: (1) Elimination of the warm bias caused by the T2m-diagnostic itself, (2) representation of uncertainty that scales with subgrid-scale spatial heterogeneity and represents a min-max temperature range, (3) probability of an otherwise unresolved temperature maximum (shallow inversion) or minimum (shallow unstable layer) below first measurement height or model level, respectively. These advancements can help to overcome identified problems in the representation of land-atmosphere coupling in NWP models by facilitating more accurate model verification studies accounting for spatial heterogeneity \citep{Remes2025}, representing unresolved features such as shallow surface-based inversions particularly in coarser models \citep{Mack2025}, and omitting (unphysical) model tuning \citep{Kaehnert2025}. The probabilistic post-processing of NWP forecasts can also mitigate the missing uncertainty due to under-dispersive ensembles in a computationally efficient way in an operational setup \citep{Ollinaho2016}. \\
At the same time, probabilistic post-processing cannot represent non-linear interactions and feedbacks between subgrid- and resolved scales, and wrongly represented surface or atmospheric temperature in the NWP model propagate to the 2 m-temperature. Therefore, future enhancements could focus on online parameterizations and incorporate more predictors, e.g., surface properties, into the underlying graphs, as well as represent the interaction between mean flow and turbulence in a setup that goes beyond a single-column model configuration.

\subsection{Copula Bayesian networks -- advancements \& limitations}
BNs combine graph and probability theory to represent causal relations uncertainty-aware.
While BNs capture interrelations between random variables with a mean-based scalar dependence measure, CBNs are based on copulas as general non-linear quantile-based dependence structure, which makes CBNs applicable in more complex scenarios. Compared to solely mean-based methods, such quantile-based methods can represent higher statistical moments, asymmetry in distributions and percentile-dependent relations in a non-Gaussian and non-linear fashion, enabling the modelling of heteroskedasticity and extreme values \citep{Cano2004}. Thus, copulas could represent a way to improve the representation of extreme events in artificial intelligence-based weather forecasting models, which are still inferior to NWP models when it comes to predicting extreme weather events \citep{Nipen2024,Zhang2025}.
The underlying modular and iteratively extendable graph structure of CBNs can be build by combining expert knowledge, physical processes and information from data. As typical for Bayesian methods, also in CBNs prior information is incorporated and belief updating is facilitated when new evidence enters \citep{Berliner1998}. \\
Limitation of the constructed CBN lies in its static nature (i.e. fixed graph structure), such that time-varying information is only indirectly present via the temperature-dependent coupling. Additional ways to incorporate time-varying information are change-point-detection \citep[e.g.][]{Xiong2015} or dynamic CBNs (DCBNs) that can consider both changes in the copulas as well as in the graph structure \citep{Eban2013}. Time-invariant information as topography or vanishing surface winds (kinematic boundary condition) cannot be modelled with copulas, which is also why constructing a similar CBN to interpolate near-surface wind speeds is not possible. \\
If copulas involved in a CBN have high entropy, the posterior density has high variance leading to high uncertainty in the inferred variable. While this property can still be seen as advantage from a Bayesian perspective, it limits practical applicability. For example, it is difficult to train a CBN for $f(SH \vert dT/dz, U)$, which would be equivalent to an online flux parameterizations, at complex sites such as Finse where counter-gradient fluxes are as likely as gradient ones, resulting in a high-variance distribution centered around zero \citep{Mack2024}. Solving this would require a coupling state-dependent dynamic CBN, which overall suffers from data scarcity, model complexity, and uncertain regime classifications \citep[e.g.][]{Kaehnert2025}.

\section{Conclusion}
Understanding land-atmosphere coupling in wintertime stable boundary layers still poses a major challenge due to the complexity of the ongoing processes resulting in a tenacious misrepresentation of cold temperature extremes in NWP models. In this study, we approach this problem from a new statistical view point by combining copula modelling, information entropy and Bayesian networks -- resulting in the following advances:

\begin{itemize}
	\item The physical process of land-atmosphere decoupling can be interpreted as an increased information loss between surface and atmosphere, that goes along with strong spatial heterogeneity, which corresponds to a horizontal information loss. To quantify information loss, we use copula entropy -- a measure for coupling uncertainty.
	\item We construct a copula Bayesian network (T2m-CBN) to couple surface and atmospheric temperature and predict near-surface temperature profiles. 
	Compared to the existing schemes used in NWP models, this new probabilistic method offers several advantages: (1) elimination of the warm bias inherent to the currently used deterministic T2m-diagnostic, (2) representation of uncertainty that scales with subgrid-scale variability, (3) probability for an otherwise unresolved temperature maximum (shallow inversion) or minimum (shallow unstable layer) below first model level. 
	The constructed parametric CBN is a standalone-product that was successfully applied to infer 2 m-temperature from observations and in post-processing of NWP model output. 
\end{itemize}
Overall, copulas and their combination with graph or information theory enable complex processes to be modeled according to their non-linear and non-Gaussian nature, opening up a wide range of possibilities in Earth system modelling.

\paragraph{Declaration of competing interest}
The authors have no relevant financial or non-financial interests to disclose.

\paragraph{Acknowledgments}
The R core team \citep{Rcoreteam2024} and the developers of the R-package "VineCopula" \citep{VineCopula2025} are acknowledged.
This work was supported by the European Research Council (project \#101116083) and is a contribution to the
strategic research initiative LATICE (Faculty of Mathematics and Natural Sciences, University of Oslo, project \#UiO/GEO103920).

\paragraph{Data availability}
The code used for the post-processing of the field data is available on GitHub (\url{https://github.com/noctiluc3nt/Reddy}).
The MEPS model output provided by the Norwegian Meteorological Institute (MET Norway) can be found in the thredds archive under \url{https://thredds.met.no/thredds/catalog/meps25epsarchive/catalog.html}. The used terrain data provided by Kartverket are available at \url{https://www.geonorge.no/}.

\appendix
\section{Stability correction functions used in T2m-diagnostic}\label{appendix:most}
The interpolation factor $s_i$ in eq. \ref{eq:t2m-interpolation}
is according to \citet{Geleyn1988} given by
\begin{align}
	s_i= \frac{1}{b_H} \cdot \left( \ln \left(1+\frac{z}{z_a} (e^{b_N}-1) \right) - \frac{z}{z_a} (b_N-b_H) \right)
\end{align}
with $b_N = \kappa/\sqrt{C_N}$ and $b_H = \kappa \, \sqrt{C_D}/C_H$ including the diffusivities
\begin{align}
	C_N &= \left( \frac{\kappa}{\ln(\frac{z_a+z_0}{z_0})} \right)^2\\
	C_D &= \frac{C_N}{f_M(Ri)} \\
	C_H &= \frac{C_N}{f_H(Ri)}
\end{align}
utilizing the stability correction functions
\begin{align}
	f_M &= 1 + \frac{10 \,Ri}{\sqrt{1+5\, Ri}} \\
	f_H &= \frac{1}{1+15\,Ri \sqrt{1+5\,Ri}}\cdot \frac{\ln (z/z_0)}{\ln(z/z_{0H})}.
\end{align}
$z_a$ is the height of the first model level, $\kappa = 0.4$ the von-Kármán constant, and the roughness length of heat is given by $z_{0H} = z_0/10$ (in MEPS).

\section{Scaling of thermal heterogeneity} \label{appendix:scaling-heterogeneity}
Based on the considered NetAtmo observations over Scandinavia, a linear relation between spatially averaged temperature $\langle T\rangle$ and subgrid-scale thermal heterogeneity $H_T$ (defined in eq. \ref{eq:thermal-heterogeneity}) in the form $H_T = a \, \langle T \rangle + b$  with
$a=-3.334\cdot10^{-6}\,dx -2.102\cdot10^{-2}$ and $b= 2.896\cdot10^{-5}\,dx + 3.085\cdot10^{-1}$ depending on the resolution $dx$ was found.

\bibliographystyle{plainnat}
\bibliography{references.bib}

\begin{thebibliography}{75}
\providecommand{\natexlab}[1]{#1}
\providecommand{\url}[1]{\texttt{#1}}
\expandafter\ifx\csname urlstyle\endcsname\relax
  \providecommand{\doi}[1]{doi: #1}\else
  \providecommand{\doi}{doi: \begingroup \urlstyle{rm}\Url}\fi

\bibitem[Aas et~al.(2009)Aas, Czado, Frigessi, and Bakken]{Aas2009}
K.~Aas, C.~Czado, A.~Frigessi, and H.~Bakken.
\newblock {Pair-copula constructions of multiple dependence}.
\newblock \emph{Insurance: Mathematics and Economics}, 44:\penalty0 182--198,
  2009.
\newblock \doi{10.1016/j.insmatheco.2007.02.001}.

\bibitem[Acevedo et~al.(2016)Acevedo, Mahrt, Puhales, Costa, Medeiros, and
  Degrazia]{Acevedo2016}
O.~C. Acevedo, L.~Mahrt, F.~S. Puhales, F.~D. Costa, L.~E. Medeiros, and G.~A.
  Degrazia.
\newblock {Contrasting structures between the decoupled and coupled states of
  the stable boundary layer}.
\newblock \emph{Q J R Meteorol Soc}, 142:\penalty0 693--702, 2016.
\newblock \doi{10.1002/qj.2693}.

\bibitem[Bauer and Czado(2016)]{Bauer2016}
A.~Bauer and C.~Czado.
\newblock {Pair-Copula Bayesian Networks}.
\newblock \emph{Journal of Computational and Graphical Statistics}, 25\penalty0
  (4):\penalty0 1248--1271, 2016.
\newblock \doi{10.1080/10618600.2015.1086355}.

\bibitem[Beljaars and Viterbo(1998)]{Beljaars1998}
A.~Beljaars and P.~Viterbo.
\newblock {The role of the boundary layer in a numerical weather prediction
  model}.
\newblock \emph{Roy. Netherlands Acad. of Arts and Sci.}, pages 287--304, 1998.

\bibitem[Beljaars et~al.(2011)Beljaars, Balsamo, Betts, Dutra, Ghelli, Köhler,
  Sandu, Serrar, and Viterbo]{Beljaars2011}
A.~Beljaars, G.~Balsamo, A.~Betts, E.~Dutra, A.~Ghelli, M.~Köhler, I.~Sandu,
  S.~Serrar, and P.~Viterbo.
\newblock {The stable boundary layer in the ECMWF model}.
\newblock \emph{ECMWF GABLS Workshop on Diurnal cycles and the stable boundary
  layer,}, 2011.

\bibitem[Berliner et~al.(1998)Berliner, Royle, Wikle, and
  Milliff]{Berliner1998}
L.~M. Berliner, J.~A. Royle, C.~K. Wikle, and R.~F. Milliff.
\newblock {Bayesian Methods in the Atmospheric Sciences}.
\newblock \emph{Bayesian Statistics}, 6, 1998.

\bibitem[Bárdossy(2006)]{Bardossy2006}
A.~Bárdossy.
\newblock {Copula-based geostatistical models for groundwater quality
  parameters}.
\newblock \emph{Water Resources Research}, 42\penalty0 (W11416), 2006.
\newblock \doi{10.1029/2005WR004754}.

\bibitem[Bárdossy and Li(2008)]{Bardossy2008}
A.~Bárdossy and J.~Li.
\newblock {Geostatistical interpolation using copulas}.
\newblock \emph{Water Resources Research}, 44\penalty0 (W07412), 2008.
\newblock \doi{10.1029/2007WR006115}.

\bibitem[Båserud et~al.(2020)Båserud, Lussana, Nipen, Seierstad, Oram, and
  Aspelien]{Baaserud2020}
L.~Båserud, C.~Lussana, T.~N. Nipen, I.~A. Seierstad, L.~Oram, and
  T.~Aspelien.
\newblock {TITAN automatic spatial quality control of meteorological in-situ
  observations}.
\newblock \emph{Adv Sci Res}, 17:\penalty0 153--163, 2020.
\newblock \doi{10.5194/asr-17-153-2020}.

\bibitem[Cano et~al.(2004)Cano, Sordo, and Gutierrez]{Cano2004}
R.~Cano, C.~Sordo, and J.~M. Gutierrez.
\newblock {Applications of Bayesian Networks in Meteorology.}
\newblock \emph{Advances in Bayesian Networks}, 2004.

\bibitem[Chen and Pollino(2012)]{Chen2012}
S.~H. Chen and C.~A. Pollino.
\newblock {Good practice in Bayesian network modelling}.
\newblock \emph{Environmental Modelling \& Software}, 37:\penalty0 134--145,
  2012.
\newblock \doi{10.1016/j.envsoft.2012.03.012}.

\bibitem[Cuxart et~al.(2006)Cuxart, Holtslag, Beare, et~al.]{Cuxart2006}
J.~Cuxart, A.~A.~M. Holtslag, R.~J. Beare, et~al.
\newblock Single-column model intercomparison for a stably stratified
  atmospheric boundary layer.
\newblock \emph{Boundary-Layer Meteorology}, 118:\penalty0 273--303, 2006.
\newblock \doi{10.1007/s10546-005-3780-1}.

\bibitem[Czado and Nagler(2022)]{Czado2022}
C.~Czado and T.~Nagler.
\newblock {Vine Copula Based Modeling}.
\newblock \emph{Annual Review of Statistics and Its Application}, 9:\penalty0
  453--477, 2022.
\newblock \doi{10.1146/annurev-statistics-040220-101153}.

\bibitem[Day et~al.(2024)Day, Svensson, Casati, Uttal, Khalsa, Bazile, Akish,
  Azouz, Ferrighi, Frank, Gallagher, God{\o}y, Hartten, Huang, Holt,
  Di~Stefano, Suomi, Mariani, Morris, O'Connor, Pirazzini, Remes, Fadeev,
  Solomon, Tjernstr\"om, and Tolstykh]{Day2024}
J.~Day, G.~Svensson, B.~Casati, T.~Uttal, S.-J. Khalsa, E.~Bazile, E.~Akish,
  N.~Azouz, L.~Ferrighi, H.~Frank, M.~Gallagher, {{\O{}}}. God{\o}y,
  L.~Hartten, L.~X. Huang, J.~Holt, M.~Di~Stefano, I.~Suomi, Z.~Mariani,
  S.~Morris, E.~O'Connor, R.~Pirazzini, T.~Remes, R.~Fadeev, A.~Solomon,
  J.~Tjernstr\"om, and M.~Tolstykh.
\newblock {The YOPP site Model Intercomparison Project (YOPPsiteMIP) phase 1:
  project overview and Arctic winter forecast evaluation}.
\newblock \emph{Geoscientific Model Development}, 17:\penalty0 5511--5543,
  2024.
\newblock \doi{10.5194/gmd-17-5511-2024}.

\bibitem[Derbyshire(1999)]{Derbyshire1998}
S.H. Derbyshire.
\newblock {Boundary-layer decoupling over cold surfaces as a physical
  boundary-instability}.
\newblock \emph{Boundary-Layer Meteorology}, 90:\penalty0 297--325, 1999.

\bibitem[Dissmann et~al.(2013)Dissmann, Brechmann, Czado, and
  Kurowicka]{Dissmann2013}
J.~Dissmann, E.C. Brechmann, C.~Czado, and D.~Kurowicka.
\newblock {Selecting and estimating regular vine copulae and application to
  financial returns}.
\newblock \emph{Computational Statistics \& Data Analysis}, 59:\penalty0
  52--69, 2013.
\newblock \doi{10.1016/j.csda.2012.08.010}.

\bibitem[Eban et~al.(2013)Eban, Rothschild, Mizrahi, Nelken, and
  Elidan]{Eban2013}
E.~Eban, G.~Rothschild, A.~Mizrahi, I.~Nelken, and G.~Elidan.
\newblock {Dynamic Copula Networks for Modeling Real-valued Time Series}.
\newblock \emph{Artificial Intelligence and Statistics}, 2013.

\bibitem[Elidan(2010)]{Elidan2010}
G.~Elidan.
\newblock {Copula Bayesian Networks}.
\newblock \emph{in: Advances in Neural Information Processing Systems 23 (NIPS
  2010)}, 23, 2010.

\bibitem[Ferguson et~al.(2012)Ferguson, Wood, and Vinukollu]{Ferguson2012}
C.~R. Ferguson, E.~F. Wood, and R.~K. Vinukollu.
\newblock {A Global Intercomparison of Modeled and Observed Land–Atmosphere
  Coupling}.
\newblock \emph{Journal of Hydrometeorology}, 13:\penalty0 749--784, 2012.
\newblock \doi{10.1175/JHM-D-11-0119.1}.

\bibitem[Frederiksen and O'Kane(2008)]{Frederiksen2008}
J.~S. Frederiksen and T.~J. O'Kane.
\newblock {Entropy, Closures and Subgrid Modeling}.
\newblock \emph{Entropy}, 10:\penalty0 635--683, 2008.
\newblock \doi{10.3390/e10040635}.

\bibitem[Frogner et~al.(2019)Frogner, Signleton, Køltzow, and
  Andrae]{Frogner2019}
I.-L. Frogner, A.~T. Signleton, M.~Ø. Køltzow, and U.~Andrae.
\newblock {Convection-permitting ensembles: Challenges related to their design
  and use}.
\newblock \emph{Q J R Meteorol Soc}, 145:\penalty0 90--106, 2019.
\newblock \doi{10.1002/qj.3525}.

\bibitem[Fréchet(1951)]{Frechet1951}
M.~Fréchet.
\newblock {Sur les tableaux de corrélation dont les marges sont données.}
\newblock \emph{Ann. Univ. Lyon Sect. A.}, 14\penalty0 (3):\penalty0 53--77,
  1951.

\bibitem[Fu et~al.(2014)Fu, Li, and Yao]{Fu2014}
Z.~Fu, N.~Li, Q. an~Yuan, and Z~Yao.
\newblock {Multi-scale entropy analysis of vertical wind variation series in
  atmospheric boundary-layer}.
\newblock \emph{Commun Nonlinear Sci Numer Simulat}, 19:\penalty0 83--91, 2014.
\newblock \doi{10.1016/j.cnsns.2013.06.026}.

\bibitem[Geleyn(1988)]{Geleyn1988}
J.-F. Geleyn.
\newblock {Interpolation of wind, temperature and humidity values from model
  levels to the height of measurement}.
\newblock \emph{Tellus}, 40A\penalty0 (4):\penalty0 347--351, 1988.

\bibitem[Genest et~al.(2006)Genest, Quessy, and Remillard]{Genest2006}
C.~Genest, J.-F. Quessy, and B.~Remillard.
\newblock {Goodness-of-fit Procedures for Copula Models Based on the
  Probability Integral Transformation}.
\newblock \emph{Scandinavian Journal of Statistics}, 33\penalty0 (2):\penalty0
  337--366, 2006.
\newblock \doi{10.1111/j.1467-9469.2006.00470.x}.

\bibitem[Gleeson et~al.(2024)Gleeson, Kurzeneva, de~Rooy, Rontu,
  et~al.]{Gleeson2024}
E.~Gleeson, E.~Kurzeneva, W.~de~Rooy, L.~Rontu, et~al.
\newblock {The Cycle 46 Configuration of the HARMONIE-AROME Forecast Model}.
\newblock \emph{Meteorology}, 3\penalty0 (4):\penalty0 354--390, 2024.
\newblock \doi{10.3390/meteorology3040018}.

\bibitem[Gouttevin et~al.(2023)Gouttevin, Vionnet, Seity, Boone, Lafaysse,
  Deliot, and Merzisen]{Gouttevin2023}
I.~Gouttevin, V.~Vionnet, Y.~Seity, A.~Boone, M.~Lafaysse, Y.~Deliot, and
  H.~Merzisen.
\newblock {To the Origin of a Wintertime Screen-Level Temperature Bias at High
  Altitude in a Kilometric NWP Model}.
\newblock \emph{Journal of Hydrometeorology}, 24:\penalty0 53--71, 2023.
\newblock \doi{10.1175/JHM-D-21-0200.1}.

\bibitem[Göber et~al.(2008)Göber, Zsótér, and Richardson]{Goeber2008}
M.~Göber, E.~Zsótér, and D.~S. Richardson.
\newblock {Could a perfect model ever satisfy a naïve forecaster? On grid box
  mean versus point verification}.
\newblock \emph{Meteorol. Appl.}, 15:\penalty0 359--365, 2008.
\newblock \doi{10.1002/met.78}.

\bibitem[Hashemi et~al.(2016)Hashemi, Khan, and Ahmed]{Hashemi2016}
S.~J. Hashemi, F.~Khan, and S.~Ahmed.
\newblock {Multivariate probabilistic safety analysis of process facilities
  using the Copula Bayesian Network model}.
\newblock \emph{Computers and Chemical Engineering}, 93:\penalty0 128--142,
  2016.
\newblock \doi{10.1016/j.compchemeng.2016.06.011}.

\bibitem[Huss and Thomas(2024)]{Huss2024}
J.-M. Huss and C.~K. Thomas.
\newblock {The impact of turbulent transport efficiency on surface vertical
  heat fluxes in the Arctic stable boundary layer predicted from similarity
  theory and machine-learning}.
\newblock \emph{Journal of the Atmospheric Sciences}, 81\penalty0 (11), 2024.
\newblock \doi{10.1175/JAS-D-24-0063.1}.

\bibitem[Joe(1996)]{Joe1996}
H.~Joe.
\newblock {Families of m-variate distributions with given margins and m(m-1)/2
  bivariate depen- dence parameters}.
\newblock \emph{in: Distributions with Fixed Marginals and Related Topics},
  1996.

\bibitem[K{\o}ltzow et~al.(2019)K{\o}ltzow, Casati, Bazile, Haiden, and
  Valkonen]{Koltzow2019}
M.~K{\o}ltzow, B.~Casati, E.~Bazile, T.~Haiden, and T.~Valkonen.
\newblock {An NWP Model Intercomparison of Surface Weather Parameters in the
  European Arctic during the Year of Polar Prediction Special Observing Period
  Northern Hemisphere 1}.
\newblock \emph{Weather and Forecasting}, 34:\penalty0 959--983, 2019.
\newblock \doi{10.1175/WAF-D-19-0003.1}.

\bibitem[Koster et~al.(2005)Koster, Guo, Dirmeyer, et~al.]{Koster2005}
R.~D. Koster, Z.~Guo, P.~A. Dirmeyer, et~al.
\newblock {GLACE: The Global Land–Atmosphere Coupling Experiment. Part I:
  Overview}.
\newblock \emph{Journal of Hydrometeorology}, 7:\penalty0 590--610, 2005.

\bibitem[Kähnert et~al.(2022)Kähnert, Sodemann, Remes, Fortelius, Bazile, and
  Esau]{Kaehnert2022}
M~Kähnert, H.~Sodemann, T.~M. Remes, C.~Fortelius, E.~Bazile, and I.~Esau.
\newblock {Spatial Variability of Nocturnal Stability Regimes in an Operational
  Weather Prediction Model}.
\newblock \emph{Boundary-Layer Meteorol}, 186:\penalty0 373--397, 2022.
\newblock \doi{10.1007/s10546-022-00762-1}.

\bibitem[Kähnert et~al.(2025)Kähnert, Sodemann, Remes, and
  Homleid]{Kaehnert2025}
M.~Kähnert, H.~Sodemann, T.~M. Remes, and M.~Homleid.
\newblock {Impact of adjustments in surface-atmosphere coupling for model
  forecasts in stable conditions}.
\newblock \emph{Weather and Forecasting}, 2025.
\newblock \doi{10.1175/WAF-D-24-0163.1}.

\bibitem[Lenderink and Holtslag(2004)]{Lenderink2004}
G.~Lenderink and A.~A.~M. Holtslag.
\newblock {An updated length-scale formulation for turbulent mixing in clear
  and cloudy boundary layers}.
\newblock \emph{Q J R Meteorol Soc}, 130:\penalty0 3405--3427, 2004.
\newblock \doi{10.1256/qj.03.117}.

\bibitem[Louis(1979)]{Louis1979}
J.~F. Louis.
\newblock {A parameteric model of vertical eddy fluxes in the atmosphere}.
\newblock \emph{Boundary-Layer Meteorol}, 17:\penalty0 187--202, 1979.

\bibitem[Ma(2020)]{Ma2020}
J.~Ma.
\newblock {Discovering Association with Copula Entropy}.
\newblock \emph{arXiv [cs.LG]}, 2020.

\bibitem[Ma(2021)]{Ma2021}
J.~Ma.
\newblock {On Thermodynamic Interpretation of Copula Entropy}.
\newblock \emph{arXiv [cs.IT]}, 2021.

\bibitem[Ma and Sun(2011)]{Ma2011}
J.~Ma and Z.~Sun.
\newblock {Mutual information is copula entropy}.
\newblock \emph{Tsinghua Science and Technology}, 16\penalty0 (1):\penalty0
  51--54, 2011.
\newblock \doi{10.1016/S1007-0214(11)70008-6}.

\bibitem[Mack et~al.(2024)Mack, Berntsen, Vercauteren, and Pirk]{Mack2024}
L.~Mack, T.~K. Berntsen, N.~Vercauteren, and N.~Pirk.
\newblock {Transfer Efficiency and Organization in Turbulent Transport over
  Alpine Tundra}.
\newblock \emph{Boundary-Layer Meteorol}, 190\penalty0 (38), 2024.
\newblock \doi{10.1007/s10546-024-00879-5}.

\bibitem[Mack et~al.(2025)Mack, Kähnert, Rauschenbach, Frank, Hasenburg, Huss,
  Jonassen, Malpas, Batrak, Remes, Pirk, and Thomas]{Mack2025}
L.~Mack, M.~Kähnert, Q.~Rauschenbach, L.~Frank, F.~H. Hasenburg, J.-M. Huss,
  M.~O. Jonassen, M.~Malpas, Y.~Batrak, T.~Remes, N.~Pirk, and C.~K. Thomas.
\newblock {Stable Boundary Layers in an Arctic Fjord‐Valley System:
  Evaluation of Temperature Profiles Observed From Fiber‐Optic Distributed
  Sensing and Comparison to Numerical Weather Prediction Systems at Different
  Resolutions}.
\newblock \emph{Journal of Geophysical Research: Atmospheres}, 130\penalty0
  (e2024JD042825), 2025.
\newblock \doi{10.1029/2024JD042825}.

\bibitem[Mahrt(2000)]{Mahrt2000}
L.~Mahrt.
\newblock {Surface heterogeneity and vertical structure of the boundary layer}.
\newblock \emph{Boundary-Layer Meteorol}, 96:\penalty0 33--62, 2000.

\bibitem[Mahrt(2014)]{Mahrt2014}
L.~Mahrt.
\newblock {Stably Stratified Atmospheric Boundary Layers}.
\newblock \emph{Annu. Rev. Fluid Mech}, 46:\penalty0 23--45, 2014.
\newblock \doi{10.1146/annurev-ﬂuid-010313-141354}.

\bibitem[Margairaz et~al.(2020{\natexlab{a}})Margairaz, Pardyjak, and
  Calaf]{Margairaz2020a}
F.~Margairaz, E.~R. Pardyjak, and M.~Calaf.
\newblock {Surface Thermal Heterogeneities and the Atmospheric Boundary Layer:
  The Thermal Heterogeneity Parameter}.
\newblock \emph{Boundary-Layer Meteorology}, 177:\penalty0 49--68,
  2020{\natexlab{a}}.
\newblock \doi{10.1007/s10546-020-00544-7}.

\bibitem[Margairaz et~al.(2020{\natexlab{b}})Margairaz, Pardyjak, and
  Calaf]{Margairaz2020b}
F.~Margairaz, E.~R. Pardyjak, and M.~Calaf.
\newblock {Surface Thermal Heterogeneities and the Atmospheric Boundary Layer:
  The Relevance of Dispersive Fluxes}.
\newblock \emph{Boundary-Layer Meteorology}, 175:\penalty0 369--395,
  2020{\natexlab{b}}.
\newblock \doi{10.1007/s10546-020-00509-w}.

\bibitem[Masson et~al.(2013)Masson, Le~Moigne, Martin, Faroux, Alias, Alkama,
  Belamari, Barbu, Boone, Bouyssel, Brousseau, Brun, Calvet, Carrer, Decharme,
  Delire, Donier, Essaouini, Gibelin, Giordani, Habets, Jidane, Kerdraon,
  Kourzeneva, Lafaysse, Lafont, Lebeaupin~Brossier, Lemonsu, Mahfouf,
  Marguinaud, Mokhtari, Morin, Pigeon, Salgado, Seity, Taillefer, Tanguy,
  Tulet, Vincendon, Vionnet, and Voldoire]{Masson2013}
V.~Masson, P.~Le~Moigne, E.~Martin, S.~Faroux, A.~Alias, R.~Alkama,
  S.~Belamari, A.~Barbu, A.~Boone, F.~Bouyssel, P.~Brousseau, E.~Brun, J.-C.
  Calvet, D.~Carrer, B.~Decharme, C.~Delire, S.~Donier, K.~Essaouini, A.-L.
  Gibelin, H.~Giordani, F.~Habets, M.~Jidane, G.~Kerdraon, E.~Kourzeneva,
  M.~Lafaysse, S.~Lafont, C.~Lebeaupin~Brossier, A.~Lemonsu, J.-F. Mahfouf,
  P.~Marguinaud, M.~Mokhtari, S.~Morin, G.~Pigeon, R.~Salgado, Y.~Seity,
  F.~Taillefer, G.~Tanguy, P.~Tulet, B.~Vincendon, V.~Vionnet, and A.~Voldoire.
\newblock {The SURFEXv7.2 land and ocean surface platform for coupled or
  offline simulation of earth surface variables and fluxes}.
\newblock \emph{Geoscientific Model Development}, 6\penalty0 (4):\penalty0
  929--960, 2013.
\newblock \doi{10.5194/gmd-6-929-2013}.

\bibitem[Monin and Obukhov(1954)]{Monin1954}
A.~S. Monin and A.~M. Obukhov.
\newblock {Basic laws of turbulent mixing in the surface layer of the
  atmosphere}.
\newblock \emph{Tr. Akad. Nauk SSSR Geophiz. Inst.}, 24\penalty0
  (151):\penalty0 163--187, 1954.

\bibitem[Nagler et~al.(2025)Nagler, Schepsmeier, Stoeber, Brechmann, and
  Graeler]{VineCopula2025}
T.~Nagler, U.~Schepsmeier, J.~Stoeber, E.~C. Brechmann, and B.~Graeler.
\newblock Vinecopula: Statistical inference of vine copulas, 2025.

\bibitem[Nieuwstadt(1984)]{Nieuwstadt1984}
F.~T.~M. Nieuwstadt.
\newblock {The Turbulent Structures of the Stable, Nocturnal Boundary Layer}.
\newblock \emph{J Atmos Sci}, 41\penalty0 (14):\penalty0 2202--2216, 1984.

\bibitem[Nipen et~al.(2020)Nipen, Seierstad, Lussana, Kristiansen, and
  Hov]{Nipen2020}
T.~N. Nipen, I.~A. Seierstad, C.~Lussana, J.~Kristiansen, and Ø. Hov.
\newblock {Adopting Citizen Observations in Operational Weather Prediction}.
\newblock \emph{Bull Americ Meteorol Soc}, 101\penalty0 (1):\penalty0 E43--E57,
  2020.
\newblock \doi{10.1175/BAMS-D-18-0237.1}.

\bibitem[Nipen et~al.(2024)Nipen, Haugen, S., Nordhagen, Salihi, Tedesco,
  Seierstad, Kristiansen, Lang, Alexe, Dramsch, Raoult, Mertes, and
  Chantry]{Nipen2024}
T.~N. Nipen, H.~H. Haugen, Ingstadm~M. S., E.~M. Nordhagen, A.~F.~S. Salihi,
  P.~Tedesco, I.~A. Seierstad, J.~Kristiansen, S.~Lang, M.~Alexe, J.~Dramsch,
  B.~Raoult, G.~Mertes, and M.~Chantry.
\newblock {Regional data-driven weather modelling with a global
  stretched-grid}.
\newblock \emph{arXiv: physics.ao-ph}, 2024.

\bibitem[O'Kane et~al.(2024)O'Kane, Harries, and Collier]{OKane2024}
T.~J. O'Kane, D.~Harries, and M.~A. Collier.
\newblock {Bayesian Structure Learning for Climate Model Evaluation}.
\newblock \emph{Journal of Advances in Modeling Earth Systems}, 16\penalty0
  (e2023MS004034), 2024.
\newblock \doi{10.1029/2023MS004034}.

\bibitem[Ollinaho et~al.(2016)Ollinaho, Lock, Leutbecher, Bechtold, Beljaars,
  Bozzo, Forbes, Haiden, Hogan, and Sandu]{Ollinaho2016}
P.~Ollinaho, S.-J. Lock, M.~Leutbecher, P.~Bechtold, A.~Beljaars, A.~Bozzo,
  R.~M. Forbes, T.~Haiden, R.~J. Hogan, and I.~Sandu.
\newblock {Towards process-level representation of model uncertainties:
  stochastically perturbed parametrizations in the ECMWF ensemble}.
\newblock \emph{Quarterly Journal of the Royal Meteorological Society},
  143\penalty0 (702):\penalty0 408--422, 2016.
\newblock \doi{10.1002/qj.2931}.

\bibitem[Peltola et~al.(2021)Peltola, Lapo, and Thomas]{Peltola2021}
O.~Peltola, K.~Lapo, and C.~K. Thomas.
\newblock {A Physics-Based Universal Indicator for Vertical Decoupling and
  Mixing Across Canopies Architectures and Dynamic Stabilities}.
\newblock \emph{Geophysical Research Letters}, 48\penalty0 (e2020GL091615),
  2021.
\newblock \doi{10.1029/2020GL091615}.

\bibitem[Pirk et~al.(2023)Pirk, Aalstad, Yilmaz, Vatne, Popp, Horvath, Bryn,
  Vollsnes, Westermann, Berntsen, Stordal, and Tallaksen]{Pirk2023}
N.~Pirk, K.~Aalstad, Y.~A. Yilmaz, A.~Vatne, A.~L. Popp, P.~Horvath, A.~Bryn,
  A.~V. Vollsnes, S.~Westermann, T.~K. Berntsen, F.~Stordal, and L.~M.
  Tallaksen.
\newblock {Snow-Vegetation-Atmosphere Interactions in Alpine Tundra}.
\newblock \emph{Biogeosciences}, 20:\penalty0 2031--2047, 2023.
\newblock \doi{10.5194/bg-20-2031-2023}.

\bibitem[Préaux et~al.(2025)Préaux, Dombrowski-Etchevers, Gouttevin, and
  Seity]{Preaux2025}
D.~Préaux, I.~Dombrowski-Etchevers, I.~Gouttevin, and Y.~Seity.
\newblock {On the proper use of temperature screen-level measurements in
  weather forecasting models over mountains}.
\newblock \emph{EGUsphere}, 2025.
\newblock \doi{10.5194/egusphere-2025-708}.

\bibitem[{R Core Team}(2024)]{Rcoreteam2024}
{R Core Team}.
\newblock \emph{R: A Language and Environment for Statistical Computing}.
\newblock R Foundation for Statistical Computing, Vienna, Austria, 2024.
\newblock URL \url{https://www.R-project.org/}.

\bibitem[Remes et~al.(2025)Remes, Køltzow, and Kähnert]{Remes2025}
T.~Remes, M.~Køltzow, and M.~Kähnert.
\newblock {Spatial variability of near-surface air temperature in the
  Copernicus Arctic regional reanalysis}.
\newblock \emph{Q J R Meteorol Soc}, 2025.
\newblock \doi{10.1002/qj.4941}.

\bibitem[Rosenblatt(1952)]{Rosenblatt1952}
M.~Rosenblatt.
\newblock {Remarks on a multivariate transformation}.
\newblock \emph{Ann. Math. Statist.}, 23\penalty0 (3):\penalty0 470--472, 1952.

\bibitem[Rudisill et~al.(2024)Rudisill, Rhoades, Xu, and Feldman]{Rudisill2024}
W.~Rudisill, A.~Rhoades, Z.~Xu, and D.~R. Feldman.
\newblock {Are Atmospheric Models Too Cold in the Mountains? The State of
  Science and Insights from the SAIL Field Campaign}.
\newblock \emph{BAMS}, 2024.
\newblock \doi{10.1175/BAMS-D-23-0082.1}.

\bibitem[Salvadori and De~Michele(2007)]{Salvadori2007}
G.~Salvadori and C.~De~Michele.
\newblock {On the Use of Copulas in Hydrology: Theory and Practice}.
\newblock \emph{Journal of Hydrological Engineering}, 12\penalty0 (4), 2007.
\newblock \doi{10.1061/(ASCE)1084-0699(2007)12:4(369}.

\bibitem[Schölzel and Friederichs(2008)]{Schoelzel2008}
C.~Schölzel and P.~Friederichs.
\newblock {Multivariate non-normally distributed random variables in climate
  research – introduction to the copula approach }.
\newblock \emph{Nonlinear Processes in Geoscience}, 15:\penalty0 761--772,
  2008.
\newblock \doi{10.5194/npg-15-761-2008}.

\bibitem[Shannon(1948)]{Shannon1948}
C.~E. Shannon.
\newblock {A Mathematical Theory of Communication}.
\newblock \emph{Bell. Syst. Tech. J.}, 27:\penalty0 379--423, 1948.

\bibitem[Sklar(1959)]{Sklar1959}
A.~Sklar.
\newblock {Fonctions de répartition à n dimensions et leurs marges}.
\newblock \emph{Publ. Inst. Stat. Univ. Paris}, 8:\penalty0 229--231, 1959.

\bibitem[Tedesco et~al.(2023)Tedesco, Lenkoski, Bloomfield, and
  Sillmann]{Tedesco2023}
P.~Tedesco, A.~Lenkoski, H.~C. Bloomfield, and J.~Sillmann.
\newblock {Gaussian copula modeling of extreme cold and weak-wind events over
  Europe conditioned on winter weather regimes}.
\newblock \emph{Environmental Research Letters}, 18\penalty0 (3), 2023.
\newblock \doi{10.1088/1748-9326/acb6aa}.

\bibitem[Tjernstroem et~al.(2021)Tjernstroem, Svensson, Magnusson,
  et~al.]{Tjernstroem2021}
M.~Tjernstroem, G.~Svensson, L.~Magnusson, et~al.
\newblock {Central Arctic weather forecasting: Confronting the ECMWF IFS with
  observations from the Arctic Ocean 2018 expedition}.
\newblock \emph{Q J R Meteorol Soc}, 147:\penalty0 1278--1299, 2021.
\newblock \doi{10.1002/qj.3971}.

\bibitem[Tootoonochi et~al.(2021)Tootoonochi, Sadegh, Haerter, Räty, Grabs,
  and Teutschbein]{Tootoonochi2021}
F.~Tootoonochi, M.~Sadegh, J.~O. Haerter, O.~Räty, T.~Grabs, and
  C.~Teutschbein.
\newblock {Copulas for hydroclimatic analysis: A practice-oriented overview}.
\newblock \emph{WIREs Water}, 2021.
\newblock \doi{10.1002/wat2.1579}.

\bibitem[van~de Wiel et~al.(2012)van~de Wiel, Moene, Jonker, Baas, Basu, Donda,
  Sun, and Holtslag]{vandeWiel2012}
B.~J.~H. van~de Wiel, A.~F. Moene, H.~J.~J. Jonker, P.~Baas, S.~Basu, J.~M.~M.
  Donda, J.~Sun, and A.~A.~M. Holtslag.
\newblock {The Minimum Wind Speed for Sustainable Turbulence in the Nocturnal
  Boundary Layer}.
\newblock \emph{Jounral of the Atmospheric Sciences}, 69:\penalty0 3116--3127,
  2012.
\newblock \doi{10.1175/JAS-D-12-0107.1}.

\bibitem[van Heerwaarden et~al.(2014)van Heerwaarden, Mellado, and
  De~Lozar]{vanHeerwaarden2014}
C.~C. van Heerwaarden, J.~P. Mellado, and A.~De~Lozar.
\newblock {Scaling Laws for the Heterogeneously Heated Free Convective Boundary
  Layer}.
\newblock \emph{Journal of the Atmospheric Sciences}, 71:\penalty0 3975--4000,
  2014.
\newblock \doi{10.1175/JAS-D-13-0383.1}.

\bibitem[Wagner et~al.(2014)Wagner, Gohm, and Rotach]{Wagner2014}
J.~S. Wagner, A.~Gohm, and M.~W. Rotach.
\newblock {The Impact of Horizontal Model Grid Resolution on the Boundary Layer
  Structure over an Idealized Valley}.
\newblock \emph{Monthly Weather Review}, 142:\penalty0 3446--3465, 2014.
\newblock \doi{10.1175/MWR-D-14-00002.1}.

\bibitem[Wesson et~al.(2003)Wesson, G., and Siqueri]{Wesson2003}
K.~H. Wesson, Katul~G. G., and M.~Siqueri.
\newblock {Quantifying organization of atmospheric turbulent eddy motion using
  nonlinear time series analysis}.
\newblock \emph{Boundary-Layer Meteorology}, 106:\penalty0 507--525, 2003.

\bibitem[Xiong et~al.(2015)Xiong, Jiang, Xu, Yu, and Guo]{Xiong2015}
L.~Xiong, C.~Jiang, C.-Y. Xu, K.~Yu, and S.~Guo.
\newblock {A framework of change-point detection for multivariate hydrological
  series}.
\newblock \emph{Water Resources Research}, 51\penalty0 (10):\penalty0
  8198--8217, 2015.
\newblock \doi{10.1002/2015WR017677}.

\bibitem[Zhang et~al.(2025)Zhang, Fischer, Zscheischler, and
  Engelke]{Zhang2025}
Z.~Zhang, E.~Fischer, J.~Zscheischler, and S.~Engelke.
\newblock {Numerical models outperform AI weather forecasts of record-breaking
  extremes}.
\newblock \emph{arXiv[physics.ao-ph]}, 2025.
\newblock \doi{10.48550/arXiv.2508.15724}.

\bibitem[Zilitinkevich et~al.(2013)Zilitinkevich, Elperin, Kleeorin,
  Rogachevskii, and Esau]{Zilitinkevich2013}
S.~S. Zilitinkevich, T.~Elperin, N.~Kleeorin, I.~Rogachevskii, and I.~Esau.
\newblock {A Hierarchy of Energy- and Flux-Budget (EFB) Turbulence Closure
  Models for Stably-Stratified Geophysical Flows}.
\newblock \emph{Boundary-Layer Meteorology}, 146:\penalty0 341--373, 2013.
\newblock \doi{10.1007/s10546-012-9768-8}.

\end{thebibliography}

\end{document}